\documentclass[11pt]{article}

\usepackage{amsmath}
\usepackage{amscd}
\usepackage{amsfonts}
\usepackage{epsfig}
\usepackage{theorem}
\usepackage{authblk}
\usepackage[titletoc]{appendix}
\usepackage[margin=0.5in]{geometry}

\newtheorem{theorem}{Theorem}[section]

\newtheorem{lemma}{Lemma}[section]

\newtheorem{definition}{Definition}[section]

\newtheorem{proposition}{Proposition}[section]

\def \C {\mathbb C}

\def \R {\mathbb R}

\def \N {\mathbb N}

\def \Z{\mathbb Z}

\def \Q {\mathbb Q}

\def \D {\mathbb D}

\title{Conditional Probability Distributions of Finite Absorbing Quantum Walks}

\author{Parker Kuklinski}

\begin{document}


\maketitle

\begin{abstract}
Quantum walks are known to have nontrivial interactions with absorbing boundaries. In particular it has been shown that an absorbing boundary in the one dimensional quantum walk partially reflects information, as observed by absorption probability computations. In this paper, we shift our sights from the local phenomena of absorption probabilities to the global behavior of finite absorbing quantum walks in one dimension. We conduct our analysis by approximating the eigenbasis of the associated absorbing quantum walk operator matrix $Q_n$ where $n$ is the lattice size. The conditional probability distributions of these finite absorbing quantum walks exhibit distinct behavior at various timescales, namely wavelike reflections for times $t=O(n)$, fractional quantum revivals for $t=O(n^2)$, and stability for $t=O(n^3)$. At the end of this paper, we demonstrate the existence of fractional quantum revivals in other sufficiently regular quantum walk systems.
\end{abstract}

\section{Introduction}

The quantum walk is a unitary analogue of the classical random walk. Where classical random walks have been used in algorithms for classical computers \cite{motwani95} \cite{dyer91} \cite{schoning02} \cite{coffman93}, quantum walks can be used in algorithms for quantum computers, providing various degrees of speedup over their classical counterparts \cite{grover96} \cite{childs09} \cite{childs03} \cite{ambainis03} \cite{ambainis07}. As opposed to the standard deviation of the random walk which grows at $O(\sqrt{t})$, the standard deviation of the quantum walk grows at $O(t)$ and the wave function has highly oscillatory behavior close to the wave fronts. The quantum walk has been studied in a variety of purely mathematical contexts \cite{kempe03} \cite{venegas-andraca12} and has also been physically implemented in a number of settings \cite{travaglione02} \cite{zahringer10}. In this paper, we will explore the asymptotic behavior of finite quantum walks with absorbing boundaries.

Quantum walks with absorbing boundaries were first studied in relation to absorption probabilities \cite{ambainis01} \cite{konno02_1}. Let $p_\infty$ be the probability that a Hadamard walk particle initialized in $|1\rangle |R\rangle$ is eventually absorbed at $|0\rangle$, and let $p_n$ be the probability that this particle is absorbed at $|0\rangle$ and not by an additional absorbing boundary at $|n\rangle$. Ambainis et.\ al.\ \cite{ambainis01} found that $p_\infty =\frac{2}{\pi}$ and $\lim _{n\rightarrow\infty}p_n=\frac{1}{\sqrt{2}}$. This paradoxical result (i.e. $\lim _{n\rightarrow\infty}p_n>p_\infty$) indicates that absorbing boundaries in the quantum walk setting partially reflect information. A sharper result conjectured by these authors and later proved by Bach and Borisov \cite{bach09} states that $p_{n+1}=\frac{1+2p_n}{2+2p_n}$. These results were extended to the three-state Grover walk \cite{wang13}, two-state quantum walks \cite{kuklinski18}, and general discrete quantum mechanical systems \cite{krovi06} \cite{kuklinski18_2}. Other authors have considered hitting times, or the mean expected time that a quantum walk particle is first observed at an absorbing boundary \cite{friedman17} \cite{varbanov08} \cite{yamasaki02}. Absorption probabilities and hitting times, however, are concerned with local behavior at an absorbing boundary as opposed to the global behavior that we wish to study. 

The quantum walk can be defined as a linear combination of translations, so it is natural to view the quantum walk operator on a finite domain as a matrix. In the case of the quantum walk with absorbing boundaries, the quantum walk operator matrix is a composition of a unitary operator with a Hermitian projection operator which projects off of the absorbing boundary locations. The asymptotic behavior of a finite quantum walk is best understood by computing the eigenbasis of the operator matrix. To compute the eigenvalues of a quantum walk matrix, we first calculate its characteristic polynomial $p_n(\lambda )$ where $n$ is the size of the domain. These characteristic polynomials satisfy a second order recursion which we exploit to compute the eigenvalues up to polynomial order. Similar techniques are used in the computation of the eigenvectors.

The eigenvalues of $Q_n$ uniformly approach two sectors of the unit circle at $O(n^{-1})$ as $n$ increases. At $t=O(n^3)$, the top eigenvalues of $Q^t_n$ begin to dominate the system. More interestingly, the minimum phase difference between the $k^\text{th}$ top eigenvalue of $Q_n$ and one of the points $\pm |a|\pm i|b|$ ($a,b$ determined by $Q_n$) is roughly $\frac{k^2}{n^2}\alpha$ for small $k$ and some constant $\alpha$. Thus, there exist times $t$ for which the top eigenvalues of $Q_n^t$ approximately align in the complex plane in various patterns. More specifically, there exists a value $\tau \in\R ^+$ such that $Q_n^{\tau n^2}$ becomes a crude approximation of the identity matrix. For sufficiently simple rational multiples $z\in\Q$ of $\tau$ (i.e. $z=p/q$ such that $p,q$ are small), the matrix $Q_n^{z\tau n^2}$ becomes a weighted sum of approximations of vector reversals and transpositions. As $z$ increases, the granularity of these approximations decreases. For an initial state $\Psi _0$ describing a delta potential directly between the absorbing boundaries, the situation becomes more visually striking. For $m\in\N$ sufficiently small, $Q_n^{m\tau /8}\Psi _0$ reproduces an approximation of $\Psi_0$. Furthermore for $p,q\in\N$ sufficiently small, $Q_n^{p\tau /8q}\Psi _0$ produces an approximation of $q$ evenly spaced delta potentials in the domain. This behavior is an approximation of fractional quantum revivals which have been observed in several quantum settings. \cite{berry01} \cite{bluhm96}

The rest of the paper is organized as follows: section 2 is dedicated to defining the finite absorbing quantum walk and its matrix representation. In section 3 we approximate the eigensystem of the one-dimensional absorbing quantum walk operator. In section 4 we use these approximations to describe the approximate fractional quantum revival behavior at $t=O(n^2)$. In section 5 we demonstrate the existence of these quantum revivals in a two dimensional absorbing quantum walk.

\section{Definitions and Methods}

To begin, we recount the quantum walk on groups as first defined by Acevedo et. al. \cite{acevedo05}. The following definitions have appeared in previous works by Kuklinski \cite{kuklinski17} \cite{kuklinski18} \cite{kuklinski18_2}.
\begin{definition}
Let $(G,\cdot )$ be a group, let $\Sigma\subset G$ where $|\Sigma |=n$, and let $U\in U(n)$ where $U(n)$ is the set of $n\times n$ unitary matrices. The \emph{quantum walk operator} $Q:\ell ^2(G\times\Sigma )\rightarrow\ell ^2(G\times\Sigma )$ corresponding to the triple $(G,\Sigma ,U)$ may be written as $Q=T(I\otimes U)$ where for $g\in G$ and $\sigma\in\Sigma$, $T:|g\rangle |\sigma\rangle\mapsto |g+\sigma\rangle |\sigma\rangle$. We denote this correspondence as $Q\leftrightarrow (G,\Sigma ,U)$. 
\end{definition}
The pair $(G,\Sigma )$ can be thought of as an undirected Cayley graph which admits loops \cite{diestel05}.

We must also define an absorption unit for quantum walks. To this end, we formally define the measurement operator. Let $b\in G\times\Sigma$ be the location of an absorption unit. The measurement operator $\Pi ^b_\text{yes}:\ell ^2(G\times\Sigma )\rightarrow\ell ^2(G\times\Sigma )$ is a projection onto $|b\rangle$ while $\Pi _\text{no}^b$ is a projection onto the the subspace spanned by elements in $(G\times\Sigma )\backslash b$. The probabilistic interpretation of quantum mechanics dictates that if we measure a state $\psi\in\ell ^2(G\times\Sigma )$ at $b$, the resulting state becomes $\Pi _\text{yes}^b\psi /\lVert\Pi _\text{yes}^b\psi\rVert$ with probability $\lVert\Pi _\text{yes}^b\psi\rVert ^2$ and $\Pi _\text{no}^b\psi /\lVert\Pi _\text{no}^b\psi\rVert$ with probability $\lVert\Pi _\text{no}^b\psi\rVert ^2$. If $B\subset G\times\Sigma$, let $\Pi _\text{no}^B$ be the composition of \emph{no} measurement projections for all $b\in B$. In this way, we can define an operator for the absorbing quantum walk.
\begin{definition}
Let $Q\leftrightarrow (G,\Sigma ,U)$ be a quantum walk operator and let $B\subset G\times\Sigma$ be a collection of absorption units. Then we say that $\Pi _\text{no}^BQ$ is the \emph{absorbing quantum walk operator} corresponding to the ordered quadruple $(G,\Sigma ,U,B)$ and we denote this correspondence as $\Pi _\text{no}^BQ\leftrightarrow (G,\Sigma ,U,B)$.
\end{definition}
We use the \emph{no} operator in our definition because if we observe the particle somewhere in $B$, then the experiment is terminated, while if the particle is not observed in $B$ (i.e. we are in the range of $\Pi _\text{no}^B$) the experiment continues. Note that we speak of the absorption units as being elements of the classical space and not as members of the corresponding orthonormal basis. 

In this paper, we are interested in the computing the probability distribution of an absorbing quantum walk $\Pi _\text{no}^BQ\leftrightarrow (G,\Sigma ,U,B)$ on $G\times\Sigma$ conditioned on the particle not being absorbed by $B$. If $\psi\in\ell ^2(G\times\Sigma )$ is an initial condition, then we are interested in calculating the following function $P_t:G\times\Sigma\rightarrow [0,1]$:
\begin{align}
P_t(x)=\frac{|\langle x|(\Pi _\text{no}^BQ)^t|\psi\rangle |^2}{\lVert (\Pi ^B_\text{no}Q)^t\psi\rVert ^2} .\
\end{align}
However, for the absorbing quantum walk it is more natural to perform analysis on the probability amplitude space before converting to the probability space, bypassing the need for renormalization at every time $t$. If $G$ is finite, we can represent the operator $\Pi _\text{no}^BQ\leftrightarrow (G,\Sigma ,U,B)$ as a $|G|\cdot |\Sigma |\times |G|\cdot |\Sigma |$ matrix.

It will often occur that we need to take large powers of $2\times 2$ matrices. Using an eigenvalue expansion, we derive:
\begin{lemma}
Let $M=\begin{bmatrix} a & b \\ c & d\end{bmatrix}$, $\lambda _\pm =\frac{1}{2}\left[ a+d\pm\sqrt{(a+d)^2-4(ad-bc)}\right]$, and $F_n=\lambda _+^n-\lambda _-^n$. Then
\begin{align}
M^n=\frac{1}{F_1}\begin{bmatrix} F_{n+1}-dF_n & bF_n \\ cF_n & F_{n+1}-aF_n\end{bmatrix}
\end{align}
and $F_{n+2}-(a+d)F_{n+1}+(ad-bc)F_n=0$.
\end{lemma}

\section{Eigensystem of the One-Dimensional Finite Absorbing Quantum Walk}

We now compute the eigensystem of the one-dimensional two-state absorbing quantum walk. The corresponding quantum walk operator can be written as $Q_n=\Pi _\text{no}^1\Pi _\text{no}^nQ\leftrightarrow (\Z ,C_1,U,\{ 1,n\} )$, which we represent as a $2n\times 2n$ unitary matrix acting as
\begin{align}
Q_n\Psi =\begin{bmatrix} 0 & U_- & 0 & \hdots & 0 \\ U_+ & 0 & U_- & ~ & ~ \\ 0 & U_+ & 0 & \ddots & ~ \\ \vdots & ~ & \ddots & \ddots & U_- \\ 0 & ~ & ~ & U_+ & 0\end{bmatrix}\begin{bmatrix} \psi _1 \\ \vdots \\ \vdots \\ \vdots \\ \psi _n\end{bmatrix} \
\end{align}
where $C_1=\{ -1,1\}$, $U=U_++U_-$, $U_+=\begin{bmatrix} a & b \\ 0 & 0\end{bmatrix}$, $U_-=\begin{bmatrix} 0 & 0 \\ -\bar{b} & \bar{a}\end{bmatrix}$, $|a|^2+|b|^2=1$, and $\psi _k=[\psi _R(k),\psi _L(k)]'$. To compute the eigenvalues of $Q_n$ we first calculate the characteristic polynomial.
\begin{proposition}
Let $p_n(\lambda )=\det\left(\lambda I-Q_n\right)$. These characteristic polynomials satisfy the following recursion:
\begin{align}
p_{n+1}(\lambda )=(\lambda ^2+1)p_n(\lambda )-|a|^2\lambda ^2p_{n-1}(\lambda )\
\end{align}
Here, $p_0(\lambda )=1$ and $p_1(\lambda )=\lambda ^2$.
\end{proposition}
{\bf Proof:} We conduct a recursive cofactor expansion on the matrix $A_n=\lambda I-Q_n$. Let $[M]_{ij}$ be the $ij$-\emph{minor} of $M$, or the matrix resulting from the deletion of the $i^\text{th}$ row and $j^\text{th}$ column. Let $B_n=[A_n]_{11}$, $C_n=[B_{n+1}]_{21}$, and $D_n=[C_n]_{12}$. Letting $A_n'=\det{A_n}$ and likewise for the other matrices, we find the following recursions:
$$A_n'=\lambda B_n',\hspace{1cm}B_n'=\lambda A_{n-1}'+bC_{n-1}',\hspace{1cm}C_n'=\bar{b}B_n'+\bar{a}D_n'\hspace{1cm}D_n'=aA_{n-1}'$$
We arrive at the result by rewriting these recursions strictly in terms of $A'$. $\hfill\Box$

While in the case of a Chebyshev recursion we are able to use a trigonometric substitution to easily facilitate locating the roots of the polynomial, this procedure will not work here due to the initial conditions \cite{gilewicz85} \cite{tran15} \cite{beraha78}. We instead must reference a set with no elementary analytic representation to describe the eigenvalues.
\begin{theorem}
Let $\Theta _n^{\pm}=\{\theta\in\C :\sin ^2{n\theta}=-y^2\sin ^2{\theta},0<\text{Re }\theta <\pi ,\pm\text{Im }\theta >0\}$ with $y=\frac{|a|}{|b|}$. Then the set $\Lambda _n$ of eigenvalues of $Q_n$ may be written as $\Lambda _n =\{ 0\}\cup\Lambda _n^+\cup\Lambda _n^-$ where
\begin{align}
\Lambda _n^\pm=\{ |a|\cos{\theta}\pm i\sqrt{1-|a|^2\cos^2{\theta}}:\theta\in\Theta _n^\pm\} .\
\end{align}
The eigenvalue $\lambda =0$ has multiplicity 2.
\end{theorem}
{\bf Proof:} Using Lemma 2.1, we can derive a closed form for the characteristic polynomial:
\begin{align}
p_n(\lambda )=\frac{\lambda ^2}{F_1(\lambda )}\left[ F_n(\lambda )-|a|^2F_{n-1}(\lambda )\right] .\
\end{align}
Here, $F_n(\lambda )=\omega _+(\lambda )^n-\omega _-(\lambda )^n$ and $\omega _\pm (\lambda )=\frac{1}{2}\left[\lambda ^2+1\pm\sqrt{(\lambda ^2+1)^2-4|a|^2\lambda ^2}\right]$. The factor of $\lambda ^2$ accounts for the eigenvalue at $\lambda =0$ of multiplicity 2. Using the substitution $\lambda _\pm (\theta )=|a|\cos{\theta}\pm i\sqrt{1-|a|^2\cos^2\theta}$, we have $F_n(\lambda _\pm (\theta ))=\pm 2i(|a|\lambda _\pm (\theta ))^n\sin n\theta$. Substituting this into the nontrivial factor and squaring gives us the equation:
$$\sin ^2n\theta =-y^2\sin^2\theta$$
Since we have squared the equation, we have doubled the number of solutions, half of which do not satisfy the original equation. We can check that only values of $\theta\in\Theta _n^\pm$ satisfy $F_n(\lambda _\pm (\theta ))-|a|^2F_{n-1}(\lambda _\pm (\theta ))=0$. $\hfill\Box$

We provide additional results to better visualize the location of the eigenvalues, the first of which gives a uniform bound on $\Lambda _n$.
\begin{proposition}
Let $S_n$ be the set defined as follows:
\begin{align}
S_n^\pm = \{ 0\}\cup\{\lambda\in\C :r(n)<\lambda <1,-\phi <\text{arg }(\pm\lambda )<\phi\}\
\end{align}
where 
\begin{align*}
r(n)^2 &= \frac{1}{2(c(n)-1)}\left[ (4|a|^2-3)-(4|a|^2-1)c\right. \\
	& \left. +\sqrt{[(2|a|+1)^2c-(4|a|^2+4|a|-1)][(2|a|-1)^2c-(4|a|^2-4|a|-1)]}\right] ,
\end{align*}
$c(n)=\left(\frac{1+|a|}{1-|a|}\right)^{\frac{1}{n-1}}$, and $e^{i\phi}=|a|+i|b|$. Then $\Lambda _n\subset (S_n^+\cup S_n^-)$.
\end{proposition}
{\bf Proof:} Following the argument of Proposition 2.2 in Kuklinski \cite{kuklinski18_2}, suppose $v$ is an eigenvector of $PU$ with eigenvalue $|\lambda |=1$ where $U$ is unitary and $P$ is a projection. Then $v$ must also be an eigenvalue of $U$. From Kuklinski \cite{kuklinski17}, the eigenvectors of the corresponding quantum walk operator with periodic boundary conditions satisfy $\lVert Pv\rVert <\lVert v\rVert$ where $P$ is the projection associated with the absorbing boundaries at $|1\rangle$ and $|n\rangle$ and $\lVert\cdot\rVert$ is the $\ell ^2$ norm. Therefore, all eigenvalues of the absorbing quantum walk operator must satisfy $|\lambda |<1$.

Suppose $p_n(\lambda _0)=0$ and $\lambda _0\ne 0$. By expanding $F_n$ in terms of $\omega _\pm$ in equation (5), we have:
$$0=\left|\frac{\omega _-(\lambda _0)-|a|^2}{\omega _+(\lambda _0)-|a|^2}\right| -\left|\frac{\omega _+(\lambda _0)}{\omega _-(\lambda _0)}\right| ^{n-1} =|f(\lambda _0)|-|g(\lambda _0)|^{n-1}.$$
If we can find a set $S\subset\D$ such that $\max _{\lambda\in S}|f(\lambda )|\le\min _{\lambda\in S}|g(\lambda )|^{n-1}$, then $\lambda _0\notin S$. Since $f(\lambda )$ is analytic on the unit disk, it obtains its maximum absolute value on the unit circle \cite{ahlfors53}, and this value is obtained at $\lambda =\pm i$ such that $|f(\lambda )|\le\frac{1+|a|}{1-|a|}$. We also find that for fixed $|\lambda |$, the function $g(\lambda )=\frac{\omega _+(\lambda )}{\omega _-(\lambda )}$ achieves its minimum absolute value at $\pm i|\lambda |$ such that $\min _{|\lambda |\le R}|g(\lambda )|= -g(iR)$. The formula for $r(n)$ follows by solving $(-g(iR))^{n-1}=\left(\frac{1+|a|}{1-|a|}\right)$ for $R$.

Through direct computation, we can prove that $|f(\lambda )|<1$ and $|g(\lambda )|>1$ on the sets $A_\pm =\{\lambda\in\C :-\phi <\text{arg }(\pm\lambda )<\phi\}$, thus completing the proof. $\hfill\Box$

By the proposition, the eigenvalues of the absorbing quantum walk operator uniformly limit to the unit circle as $n$ increases. Using the identity $\lim _{n\rightarrow\infty}n(x^{1/n}-1)=\log{x}$, we can show that the following approximation on our bound holds:
$$r(n)=1-\frac{|a|^2}{n}\log\frac{1+|a|}{1-|a|}+O(n^{-2})$$
Furthermore, the eigenvalues are restricted to two sectors of the unit disk symmetric about the real axis; as $|a|$ increases these sectors become larger. These features of the eigenvalues can be observed in Figure 1.

We now make asymptotic approximations on the eigenvalues of $Q_n$. First we write an asymptotic description of elements of $\Theta _n$, which may be verified via direct computation.
\begin{lemma}
Consider the set $\Theta _n^\pm$ from Theorem 4.1 with elements $\theta _{k,n}^\pm\in\Theta _n^\pm$ for $k\in\{ 1,...,n-1\}$. For fixed $k$, let $x=\pi k$ and $y=\frac{|a|}{|b|}$ such that:
\begin{align}
\theta ^\pm _{k,n}=\frac{x}{n}\pm\frac{ixy}{n^2}-\frac{xy^2}{n^3}\mp\frac{ixy}{n^4}\left( iy^2+\frac{x^2}{6}+\frac{x^2y^2}{6}\right) +O(n^{-5}).
\end{align}
For $\alpha =\frac{k}{n}$ fixed, we have:
\begin{align}
\theta _{\alpha n,n}^\pm=\pi\alpha\pm\frac{i}{n}\sinh ^{-1}\left(y\sin\pi\alpha\right)+O(n^{-2}).
\end{align}
\end{lemma}
{\bf Proof:} Recall that we are solving the equation $\sin^2 n\theta =-y^2\sin ^2\theta$, where the roots of positive imaginary real part belong to $\Theta _n^+$ and the roots of negative imaginary real part belong to $\Theta _N^-$. Otherwise, we are solving two equations $\sin{n\theta}=\pm iy\sin\theta$ (here, $\pm$ is independent of $\Theta _n^\pm$) and separating solutions into the sets $\Theta _n^\pm$ afterwards. By using the representation $\theta =\frac{\theta _1}{n}+O(n^{-1})$, we find that $\sin\theta _1+O(n^{-1})=O(n^{-1})$, and therefore $\theta _1=\pi k$ for some $k\in\Z$. By further considering an $N^\text{th}$ order approximation of $\theta =\sum _{j=1}^N\theta _jn^{-j}+O(n^{-(N+1)})$ and using an angle sum identity, we have
$$(-1)^k\sin\left(\sum _{j=1}^{N-1}\frac{\theta _{j+1}}{n^{j}}\right) +O(n^{-N})=\pm iy\sin\left(\sum _{j=1}^{N-1}\frac{\theta _j}{n^j}\right) +O(n^{-N})$$
Each choice of $k$ corresponds to two root approximations dictated by the $\pm$ sign on the right hand side. If we eliminate the factor of $(-1)^k$ on the left side, then the two root approximations indexed by the $\pm$ sign correspond to roots with positive and negative imaginary part respectively, and thus are in direct correspondence with the sets $\Theta _n^\pm$. By choosing $N=4$ and using a Taylor expansion of $\sin\theta$, we can write:
$$\frac{\theta _2}{n}+\frac{\theta _3}{n^2}+\frac{\theta _4}{n^3}-\frac{1}{6}\left(\frac{\theta _2}{n}\right) ^3+O(n^{-4})=\pm iy\left(\frac{\theta _1}{n}+\frac{\theta _2}{n^2}+\frac{\theta _3}{n^3}-\frac{1}{6}\left(\frac{\theta _1}{n}\right) ^3\right) +O(n^{-4})$$
By equating like factors of $n^{-j}$, we have the following system of three equations:
$$\theta _2=\pm iy\theta _1,\hspace{1cm}\theta _3=\pm iy\theta _2,\hspace{1cm}\theta _4-\frac{\theta _2^3}{6}=\pm iy\left(\theta _3-\frac{\theta _1^3}{6}\right)$$
Solving this system gives the first result. If at the outset we instead fix $\alpha =\frac{k}{n}$, we instead have the expansion $\theta =\pi\alpha +\frac{\theta _1}{n}+O(n^{-2})$, thus giving the equation
$$\sin\left(\frac{\theta _1}{n}\right)=\pm iy\sin\pi\alpha$$
Solving this for $\theta _1$ gives us the second result.
$\hfill\Box$

For the remainder of the paper we let $y=|a|/|b|$. This lemma allows us to index our approximations of the nonzero eigenvalues of $Q_n$ as $\lambda _{k,n}^\pm =\lambda _\pm (\theta ^\pm _{k,n})$. It should be noted that $\theta =\pi +\theta _{k,n}^\pm$ is also a solution to $\sin n\theta =\pm iy\sin\theta$. These solutions, when passed through the function $\lambda _\pm (\theta )$, represent eigenvalues which converge to $-e^{\pm i\phi}$. Since the characteristic polynomial $p_n(\lambda )$ in equation (3) is an even function, we will omit mention of these solutions without loss of generality. When presenting asymptotic characterizations of these eigenvalues, we distinguish between the convergence of the $k^\text{th}$ eigenvalue to $\pm e^{\pm i\phi}$ (as given by $\lambda _{k,n}$)  and the uniform convergence of the collection of eigenvalues to arcs of the unit circle as described by Proposition 3.2 (these eigenvalues represented as $\lambda _{\alpha n,n}$). We summarize this in the following proposition:
\begin{theorem}
For fixed $k$, we write $\lambda _{k,n}^\pm\in\Lambda _n$ from Theorem 4.1 as follows:
\begin{align}
\lambda _{k,n}^\pm =e^{\pm i\phi}\left( 1\mp\frac{ix^2y}{2n^2} -\frac{x^2y^2}{n^3}-\frac{x^2y}{n^4}\left[\pm\frac{3}{2}y^2+\frac{x^2}{24}\left( 3y\pm i(3y^2+1)\right)\right]\right) +O(n^{-5}).
\end{align}
For $|\alpha |<1$ with $\alpha n\in\Z$, we can write
\begin{align}
\lambda _{\alpha n,n}^\pm=e^{\pm if(\alpha )}\left( 1-\frac{1}{n}\frac{|a|\sin\pi\alpha}{\sqrt{1-|a|^2\cos^2\pi\alpha}}\sinh^{-1}\left(\frac{|a|}{|b|}\sin\pi\alpha\right)\right) +O(n^{-2})
\end{align}
where $e^{\pm if(\alpha )}=|a|\cos\pi\alpha\pm i\sqrt{1-|a|^2\cos^2\pi\alpha}$.
\end{theorem}
{\bf Proof:} We can formally write $\theta _{k,n}^\pm =\theta _\pm\left(\frac{1}{n}\right)$ where the coefficient of $n^{-m}$ in equation (8) is represented by $\theta ^{(m)}_\pm (0)/m!$. In this way, we can write $\lambda _{k,n}^\pm =\lambda _\pm\left(\theta _\pm\left(\frac{1}{n}\right)\right)$. If we abuse notation and define, for the moment, $\lambda =\lambda _\pm (0)$, $\theta =\theta _\pm (0)$, and likewise for higher derivatives, then by the chain rule we have:
\begin{align}
\lambda _{k,n}^\pm &= \lambda +\frac{\theta '\lambda '}{n}+\frac{\theta ''\lambda '+(\theta ')^2\lambda ''}{2n^2}+\frac{\theta '''\lambda '+3\theta '\theta ''\lambda ''+(\theta ')^3\lambda '''}{6n^3} \\ \nonumber
	&+\frac{\theta ''''\lambda '+(4\theta '\theta '''+3(\theta '')^2)\lambda '' +6(\theta ')^2\theta ''\lambda '''+(\theta ')^4\lambda ''''}{24n^4}+O(n^{-5})
\end{align}
Lemma 3.1 gives us the ``derivatives" $\theta ^{(m)}$, while implicit differentiation of the equation $\lambda _\pm (\theta )^2-2|a|\lambda _\pm (\theta )\cos\theta +1=0$ with respect to $\theta$ gives us the derivatives $\lambda ^{(m)}$. Since $\lambda (\theta )$ is an even function, we need only consider the even derivatives:
$$\lambda =e^{\pm i\phi},\hspace{1cm}\lambda ''=\pm iye^{\pm i\phi},\hspace{1cm}\lambda ^{(4)}=-ye^{\pm i\phi}\left[ 3y\pm i(3y^2+1)\right]$$
Making these substitutions gives us the first result. To arrive at the second, we use a Taylor approximation of $\lambda _\pm(\theta )$ at $\theta =\pi\alpha$ instead of at $\theta =0$ as in the previous case. Combining equation (12) with the approximations in Lemma 3.1 gives us the second result. $\hfill\Box$

\begin{figure}
\begin{center}\includegraphics[scale=0.65]{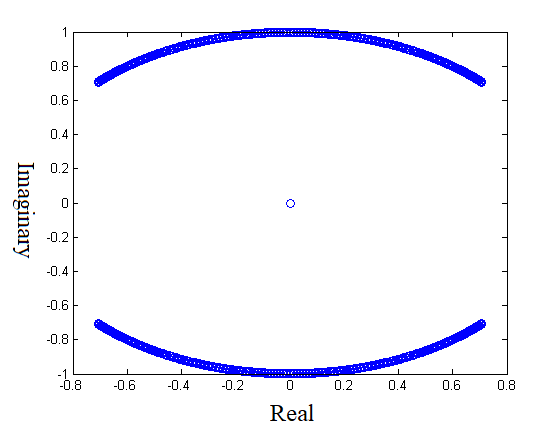}\includegraphics[scale=0.65]{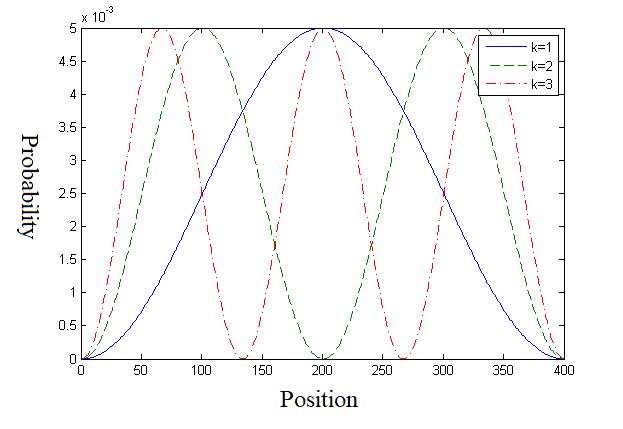}\end{center}
\caption{Plots of the eigenbasis for $n=200$ and $a=\frac{1}{\sqrt{2}}$ (\emph{Left}) Location of eigenvalues (\emph{Right}) Plots of $|v_{k,n}^\pm |^2$}
\end{figure}

With the results from Theorem 3.2, we can prove that the absorbing quantum walk reaches a steady state at time $t=O(n^3)$.
\begin{proposition}
Let $\epsilon >0$ be fixed. There exists $T=\frac{\log (1/\epsilon )}{\pi ^2y^2(k^2-1)}n^3 +O(n^2)$ such that for all $t>T$, we have
$$\left(\frac{|\lambda _{k,n}^\pm |}{|\lambda _{1,n}^\pm |}\right) ^t<\epsilon$$
\end{proposition}
This proposition shows that the magnitude of the ratio between the largest and $k^\text{th}$ largest eigenvalues of $Q_n^t$ becomes arbitrarily small at times  $t=O(n^3)$. This implies that the largest eigenmode becomes dominant at this timescale. Compare this behavior to the classical random walk which reaches a stable distribution at $t=O(n^2)$.

Using similar techniques we can compute entries of the corresponding eigenvectors.
\begin{theorem}
Let $v^\pm _{k,n}=[r^\pm _{k,n,1},l^\pm _{k,n,1},...,r^\pm _{k,n,n},l^\pm _{k,n,n}]'$ be the eigenvector of $Q_n$ corresponding to eigenvalue $\lambda ^\pm _{k,n}$. By letting $j=n\beta$ for $|\beta |<1$ fixed, we can write these eigenvectors as
\begin{align}
\begin{bmatrix} r_{k,n,\beta n}^\pm \\ l_{k,n,\beta n}^\pm\end{bmatrix}=\left(\frac{a}{|a|}\right) ^{n\beta}\left(\begin{bmatrix}\bar{a}b \\ \pm i|a||b|\end{bmatrix}\sin\pi k\beta +\frac{\pi k}{n}\begin{bmatrix} \bar{a}b(\pm i\beta y-1) \\ |a|^2(\beta -1)\end{bmatrix}\cos\pi k\beta+O(n^{-2})\right)
\end{align}
\end{theorem}
{\bf Proof:} We find that the entries of $v^\pm _{k,n}$ satisfy a matrix multiplication recursion:
$$\begin{bmatrix} r^\pm _{k,n,j+1} \\ l^\pm _{k,n,j+1}\end{bmatrix} =\frac{1}{\lambda ^\pm _{k,n}}\begin{bmatrix} a & b \\ \frac{a\bar{b}}{\bar{a}} & \frac{(\lambda ^\pm _{k,n})^2+|b|^2}{\bar{a}}\end{bmatrix}\begin{bmatrix} r^\pm _{k,n,j} \\ l^\pm _{k,n,j}\end{bmatrix}$$
Using Lemma 2.1 gives us the expression:
$$\begin{bmatrix} r_{k,n,j}^\pm \\ l_{k,n,j}^\pm\end{bmatrix}=\frac{1}{(\bar{a}\lambda )^jF_1}\begin{bmatrix} F_j-(\lambda ^2+|b|^2)F_{j-1} & \bar{a}bF_{j-1} \\ a\bar{b}F_{j-1} & F_j-|a|^2F_{j-1}\end{bmatrix}\begin{bmatrix} r_{k,n,1}^\pm \\ l_{k,n,1}^\pm\end{bmatrix}$$
Since the nonzero eigenvalues must satisfy $r_{k,n,1}^\pm =0$, we have:
$$\begin{bmatrix} r_{k,n,j}^\pm \\ l_{k,n,j}^\pm\end{bmatrix}=\frac{1}{(\bar{a}\lambda )^jF_1}\begin{bmatrix}\bar{a}bF_{j-1} \\ F_j-|a|^2F_{j-1}\end{bmatrix}$$
By making the substitutions $j=n\beta$ and $\lambda (\theta )$, and using equations (8) and (10) to make approximations for $\theta$ and $\lambda$ respectively, we arrive at the result. $\hfill\Box$

On the right side of Figure 1, we see that the top eigenvectors are are approximations of sine waves up to phase, as described by Theorem 3.3. We pause for a moment to consider these results as an analogy to the partical in an infinite well \cite{griffiths82}. The approximately sinusoidal top eigenvectors of the presently defined absorbing quantum walk operator $Q_n$ are in agreement with eigenfunctions of the particle in an infinite potential well. Moreover, the phase difference between the $k^\text{th}$ top eigenvalue of $Q_n$ and one of the points $\pm |a|\pm i|b|$ is proportional to $k^2$ for $n$ sufficiently large. This same energy spacing is present in the infinite potential well particle and leads to the revival behavior of the following section.

\section{Fractional Quantum Revivals}

Previous studies have illustrated that the quantum walk has wave-like behavior for $t=O(n)$ \cite{ambainis01}, and in particular that quantum walks appear to partially reflect off absorbing boundaries \cite{oliveira06} \cite{kuklinski18}. In the previous section, we showed that the absorbing quantum walk reaches a stable state for $t=O(n^3)$. At the intermediate timescale $t=O(n^2)$, the absorbing quantum walk exhibits fractional quantum revivals. \cite{berry01} \cite{bluhm96}

First, let us consider an approximation of $(\lambda ^\pm _{k,n}e^{\mp i\phi})^{\tau n^2}$ for a specific value of $\tau$:
\begin{proposition}
For $\tau =4/(\pi y)$, the following result holds:
\begin{align}
(\lambda ^\pm _{k,n}e^{\mp i\phi})^{\tau n^2}=1+O(n^{-1})
\end{align}
\end{proposition}
{\bf Proof:} Using Theorem 3.2, we can take a logarithm to find:
\begin{align*}
\log\left[ (\lambda ^\pm _{k,n}e^{\mp i\phi})^{\tau n^2}\right] &= (\tau n^2)\log (\lambda ^\pm _{k,n}e^{\mp i\phi}) \\
	&= \tau n^2\left(\mp i\frac{(\pi k)^2y}{2n^2}+O(n^{-3})\right) \\
	&= \mp i\tau y\frac{(\pi k)^2}{2}+O(n^{-1})
\end{align*}
Letting $\tau =4/(\pi y)$ results in the main component becoming $\pm 2\pi i k^2$ and the result immediately follows. $\hfill\Box$

For the remainder of the paper, we let $\tau =4/(\pi y)$. This result shows that the top few eigenvalues of $Q _n^{\tau n^2}$ are approximately equal to one. However, these approximations are only suitable for fixed $k$; for $t=O(n^2)$ numerics show us that for $\epsilon >0$ fixed there are $O(\sqrt{n})$ eigenvalues with $|(\lambda _{k,n}^\pm )^t|>\epsilon$. Thus we would like new approximation for the eigenvalues $\lambda _{\beta\sqrt{n},n}^\pm$ where $\beta$ is fixed. We write an extension of Lemma 3.1:
\begin{lemma}
Consider the set $\Theta _n^{\pm}$ from Theorem 3.1 with elements $\theta _{k,n}\in\Theta ^\pm _n$ for $k\in\{ 1,...,n-1\}$. For $\beta =\frac{k}{\sqrt{n}}$ fixed, we have:
\begin{align}
\theta _{\beta\sqrt{n},n}^\pm =\frac{x}{\sqrt{n}}\pm\frac{ixy}{n^{3/2}}\pm\frac{ixy}{n^{5/2}}\left[ \pm iy-\frac{x^2}{6}(1+y^2)\right] +O(n^{-7/2})
\end{align}
where $x=\pi\beta$.
\end{lemma}
{\bf Proof:} We follow the argument of Lemma 3.1 and fix $\beta =\frac{k}{\sqrt{n}}$. This allows us to posit an asymptotic expansion $\theta =\sum _{j=1}^N\frac{\theta _j}{n^{j-1/2}}+O(n^{-(N+1/2)})$, giving us an equation:
$$\sin\left(\frac{\theta _2}{\sqrt{n}}+\frac{\theta _3}{n^{3/2}}\right) +O(n^{-5/2})=\pm iy\sin\left(\frac{\theta _1}{\sqrt{n}}+\frac{\theta _2}{n^{3/2}}\right) +O(n^{-5/2})$$
The first term becomes $\theta _1=\pi\beta$. By using a Taylor series expansion of $\sin\theta$ and matching like powers of $n^{-j}$, we have:
$$\theta _2=\pm iy\theta _1,\hspace{1cm}\theta _3-\frac{\theta _2^3}{6}=\pm iy\left(\theta _2-\frac{\theta _1^3}{6}\right)$$
Solving this system gives the result. $\hfill\Box$

Using Lemma 4.1, we can find the eigenvalues at this new timescale.
\begin{proposition}
For fixed $\beta =\frac{k}{\sqrt{n}}$, we write $\lambda _{\beta\sqrt{n},n}^\pm\in\Lambda _n$ from Theorem 4.1 as follows:
\begin{align}
\lambda _{k,n}^\pm =e^{\pm i\phi}\left( 1\pm\frac{ix^2y}{2n}-\frac{x^2y}{n^2}\left(y+\frac{x^2}{24}\left[ 3y\pm i(3y^2+1)\right]\right)\right) +O(n^{-3})
\end{align}
\end{proposition}
{\bf Proof:} We can formally write $\theta _{\beta\sqrt{n},n}^\pm =\varphi _\pm\left(\frac{1}{\sqrt{n}}\right)$ where the coefficient of $n^{-(m-1/2)}$ is represented by $\varphi ^{(2m-1)}(0)/(2m-1)!$. This allows us to write $\lambda _{\beta\sqrt{n},n}^\pm =\lambda _\pm\left(\varphi _\pm\left(\frac{1}{\sqrt{n}}\right)\right)$. By using the same chain rule argument but noting that $\varphi ^{(2m+1)}=0$ and $\lambda ^{(2m)}=0$, we have:
$$\lambda _{\beta\sqrt{n},n}^\pm =\lambda +\frac{(\varphi ')^2\lambda ''}{2n}+\frac{4\varphi '\varphi '''\lambda ''+(\varphi ')^4\lambda ''''}{24n^2}+O(n^{-3})$$
Making the proper substitutions gives us the result. $\hfill\Box$

With this new representation of the eigenvalues, we can consider a more detailed approximation of the relevant eigenvalues at the timescale $t=O(n^2)$. 
\begin{theorem}
For $\beta =k/\sqrt{n}$ fixed, we have:
\begin{align}
(\lambda ^\pm _{\beta\sqrt{n},n}e^{\mp i\phi})^{\tau n^2+\rho n}=e^{-x^2y^2\tau}\exp\left[\pm\frac{ix^2y}{2}\left(\rho -\frac{x^2\tau}{12}(3y^2+1)\right)\right] +O(n^{-1})
\end{align}
\end{theorem}
{\bf Proof:} Taking a logarithm of the left hand side of equation (17) and using the approximation from proposition 4.2, we have:
\begin{align*}
\log\left[ (\lambda ^\pm _{\beta\sqrt{n},n}e^{\mp i\phi})^{\tau n^2+\rho n}\right] &= (\tau n^2+\rho n)\log (\lambda ^\pm _{\beta\sqrt{n},n}e^{\mp i\phi}) \\
	&= (\tau n^2+\rho n)\log\left( 1\pm\frac{ix^2y}{2n}-\frac{x^2y}{n^2}\left(y+\frac{x^2}{24}\left[ 3y\pm i(3y^2+1)\right]\right) +O(n^{-3})\right) \\
	&= (\tau n^2+\rho n)\left[\left(\pm\frac{ix^2y}{2n}-\frac{x^2y}{n^2}\left(y+\frac{x^2}{24}\left[ 3y\pm i(3y^2+1)\right]\right)\right) -\frac{1}{2}\left(\pm\frac{ix^2y}{2n}\right) ^2\right] +O(n^{-1}) \\
	&= \pm\frac{ix^2y}{2}\tau n-x^2y^2\tau\pm\frac{ix^2y}{2}\left[\rho -\frac{x^2\tau}{12}(3y^2+1)\right] +O(n^{-1})
\end{align*}
The result follows from noting that $\pm ix^2y\tau n/2$ is always an integer. $\hfill\Box$

This theorem gives us a better perspective on the eigenvalues of $Q_n^t$ for $t=O(n^2)$. The absolute value of equation (17) decreases at $O(e^{-\beta ^2})$ for large $\beta$, and the phase is a quartic function of $\beta$. More importantly, equation (17) shows that the top $O(\sqrt{n})$ eigenvalues of $Q_n$ are significant for $t=O(n^2)$ and roughly align in the complex plane on the real interval $[0,1]$. The remaining eigenvalues become negligible at this scale.

One reason we introduce the linear term $\rho n$ to the exponent is to better ``align" the eigenvalues in the complex plane. It is an interpretive task to define alignment, however one particularly compelling definition is to consider alignment as an entropy minimization task. Recall that if $p:\R\rightarrow\R$ is a probability distribution function, its \emph{Shannon entropy} (not to be confused with Von Neumann entropy of a density matrix) is defined as $H[p]=-\int _{-\infty}^\infty p(x)\log{p(x)}dx$. Loosely speaking, entropy is a measure of uncertainty in the outcome of a probabilistic process. For instance, if $p(x)=\delta _{x_0}(x)$, then $H[p]=0$, otherwise $p$ describes a deterministic quantity and there is no uncertainty in the corresponding outcome. Alternatively, if the domain of $p$ is an interval on the real line, $H[p]$ is maximized when $p$ is the uniform distribution, or all possible outcomes are equally likely. Suppose we begin our absorbing quantum walk with the initial state $\psi _0=|\frac{n}{2}\rangle |R\rangle$. Notice that $H[|\psi _0|^2]=0$. We claim that at time $t=\tau n^2+O(n)$, $Q_n^t$ in some sense approximates the identity matrix; the fitness of this approximation can be estimated by computing the entropy quantity $H[|Q_n^t\psi _0|^2]$. If $Q_n^t$ is a good approximation of the identity matrix, we expect this entropy quantity to be small, and subsequently for the eigenvalues of $Q_n^t$ to ``align" in some capacity. This method of using entropy to gauge quantum revivals was explored previously in Romera and de Los Santos \cite{romera07}. We explore this concept visually and computationally in the remainder of the section.

\begin{figure}
\begin{center}\includegraphics[scale=0.65]{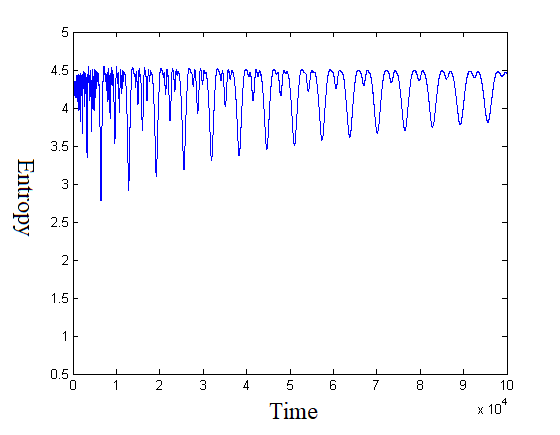}\includegraphics[scale=0.65]{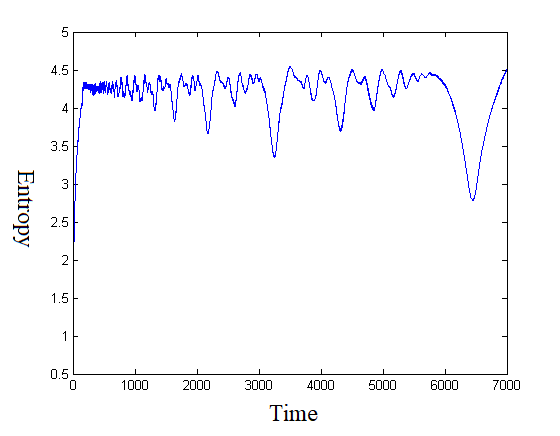}\end{center}
\caption{Plots of entropy over time for $H[|Q_n^t\psi _0|^2]$ with $n=200$, $y=1$, and $\psi _0=|100\rangle |R\rangle$.}
\end{figure}

\begin{figure}
\begin{center}
\includegraphics[scale=0.12]{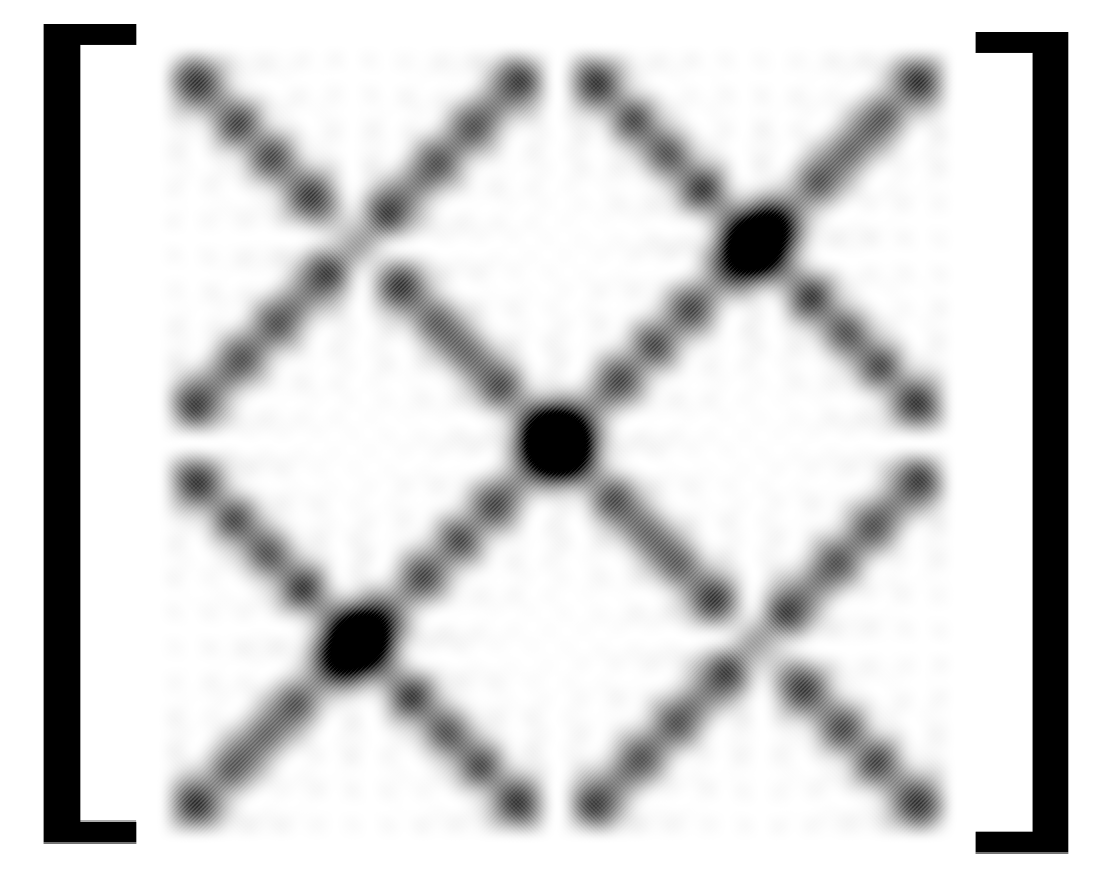}\hspace{0.05cm}\includegraphics[scale=0.12]{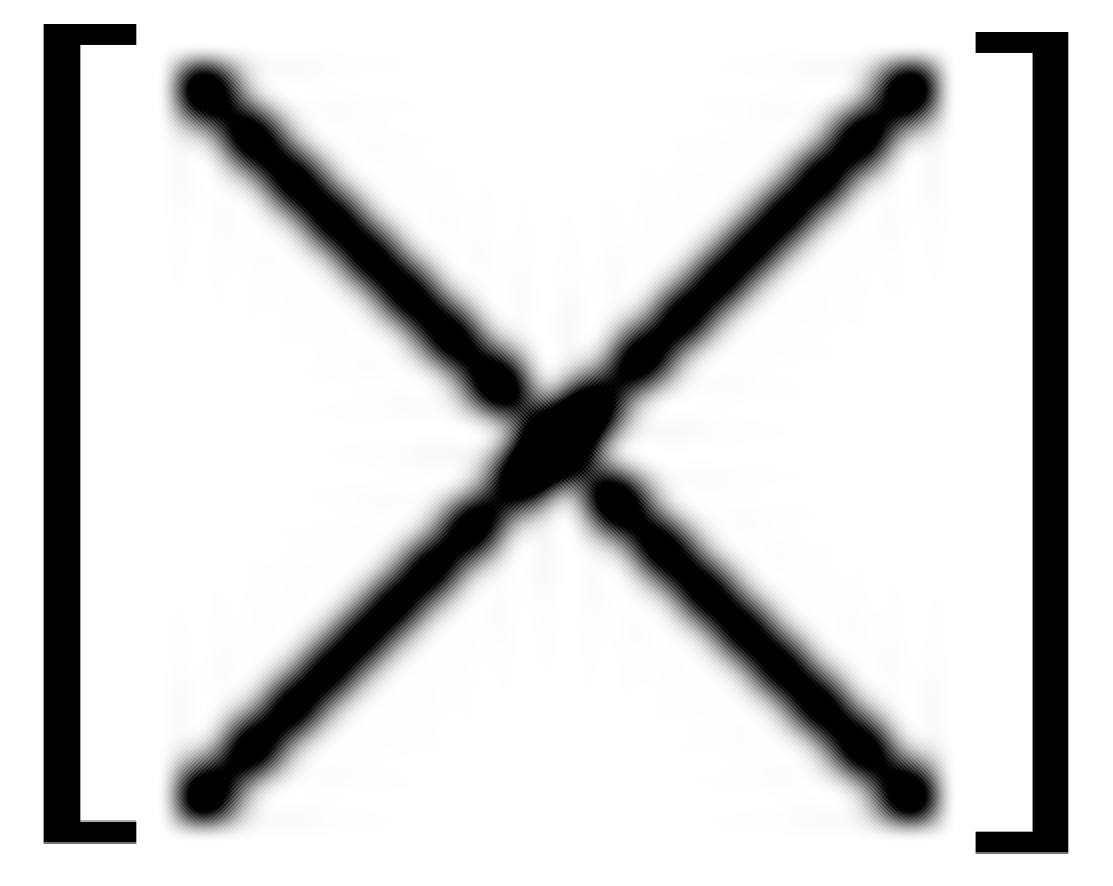}\hspace{0.05cm}\includegraphics[scale=0.12]{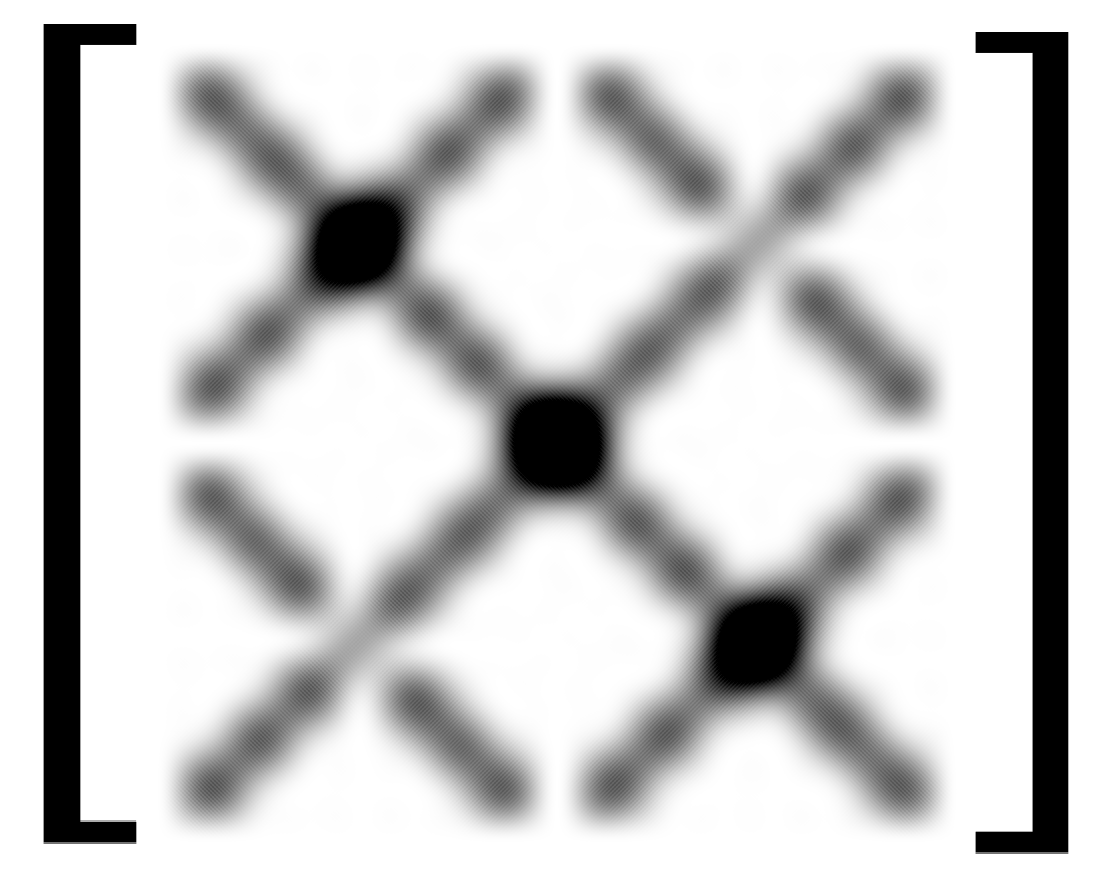}\hspace{0.05cm}\includegraphics[scale=0.12]{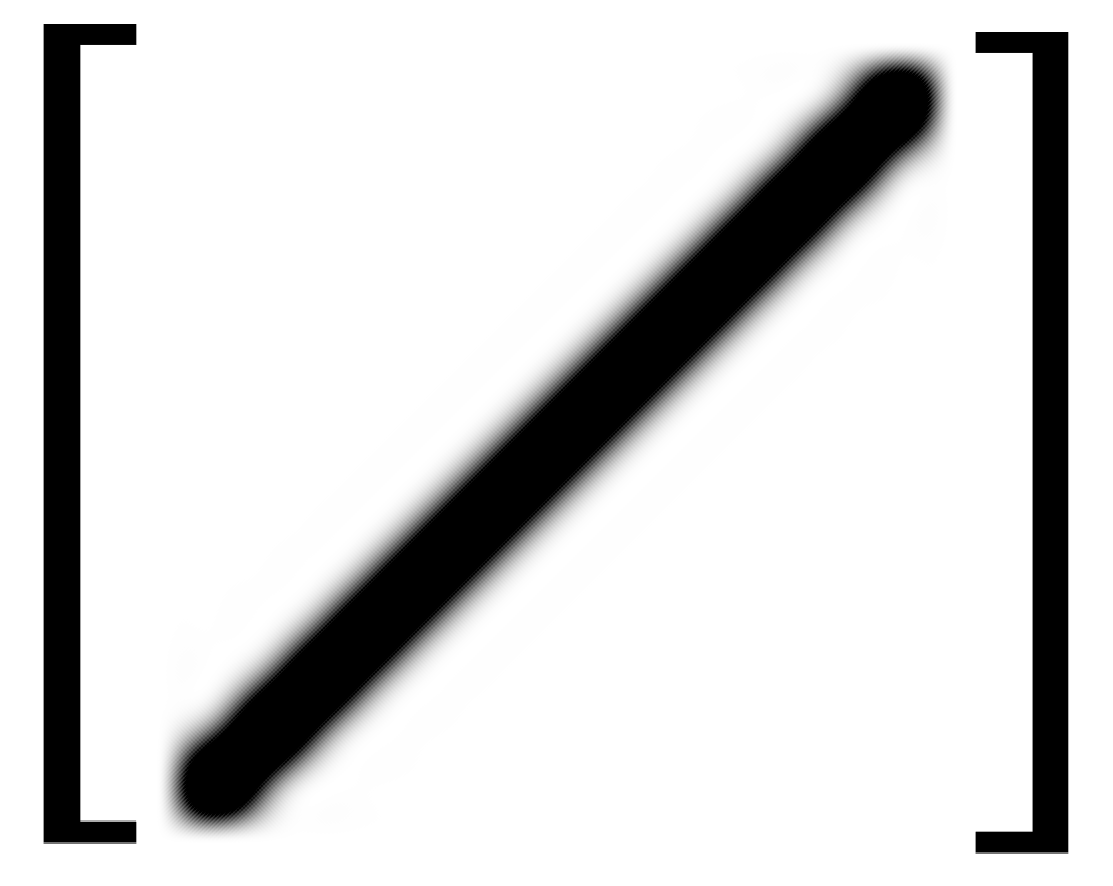}

\vspace{0.5cm}

\includegraphics[scale=0.12]{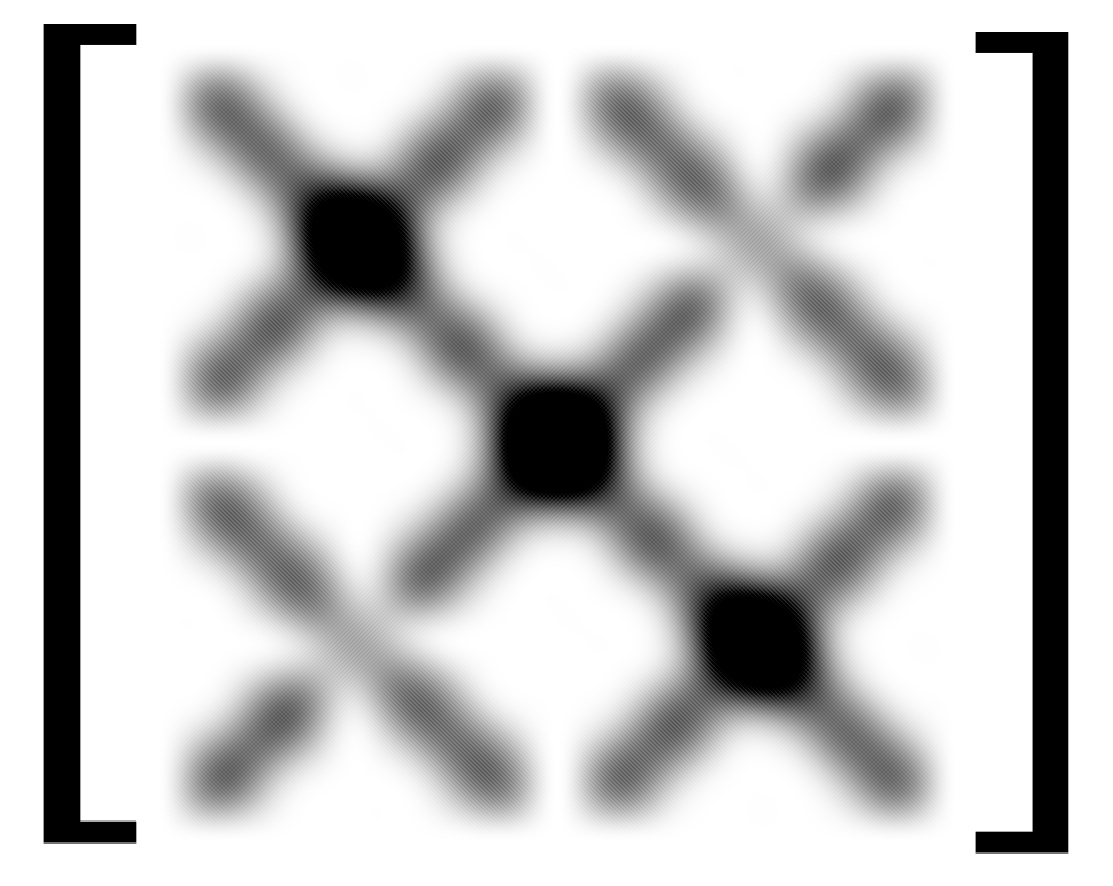}\hspace{0.05cm}\includegraphics[scale=0.12]{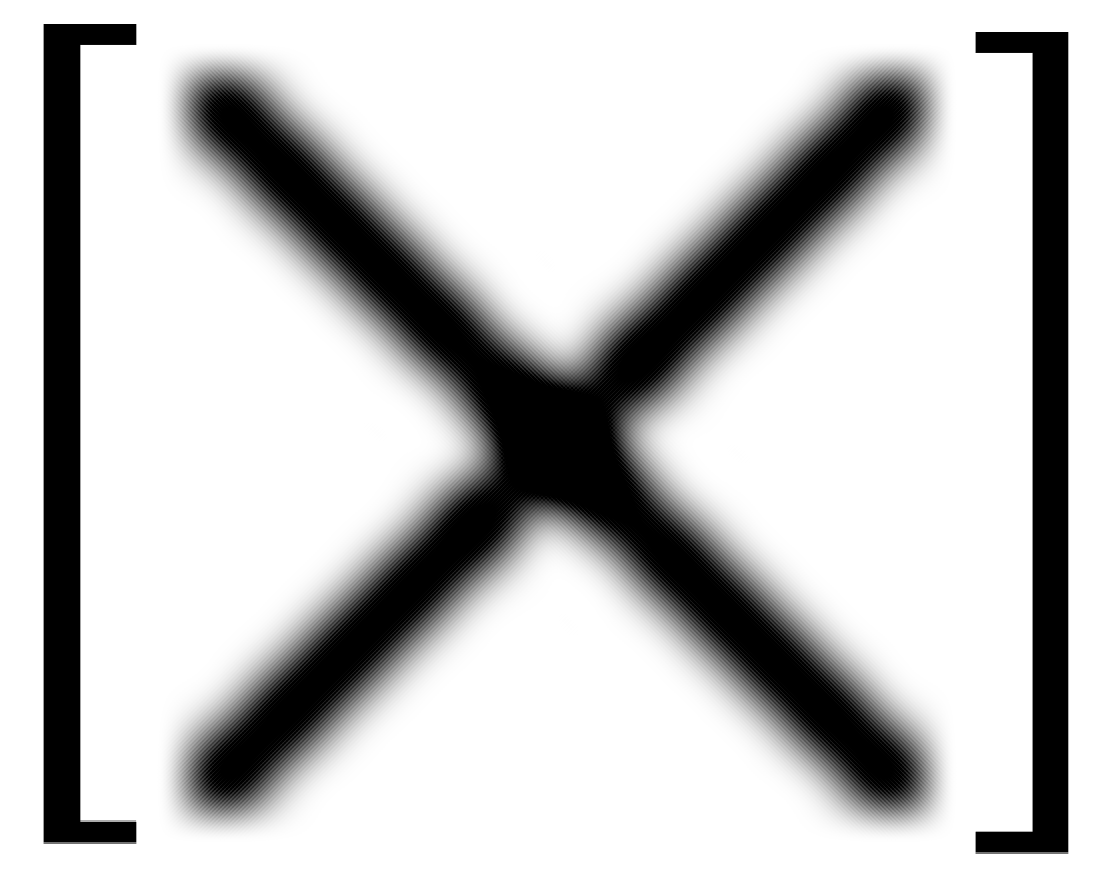}\hspace{0.05cm}\includegraphics[scale=0.12]{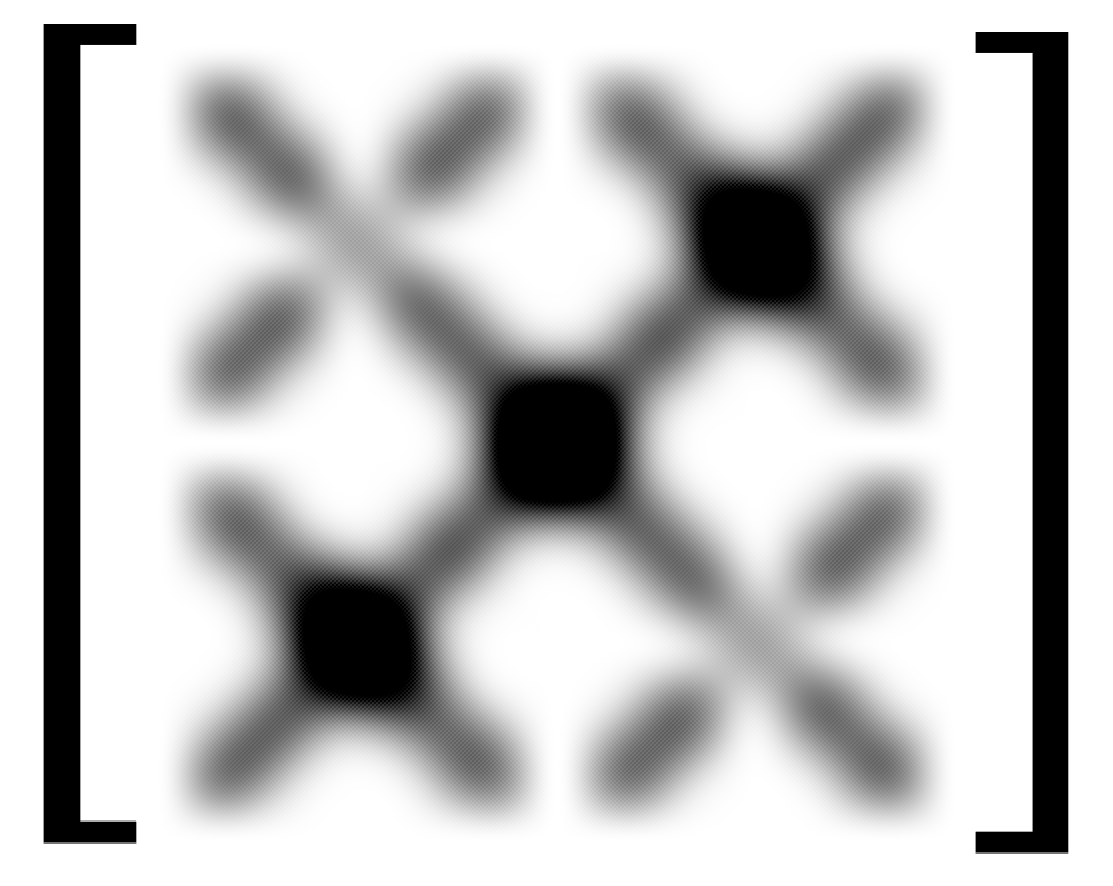}\hspace{0.05cm}\includegraphics[scale=0.12]{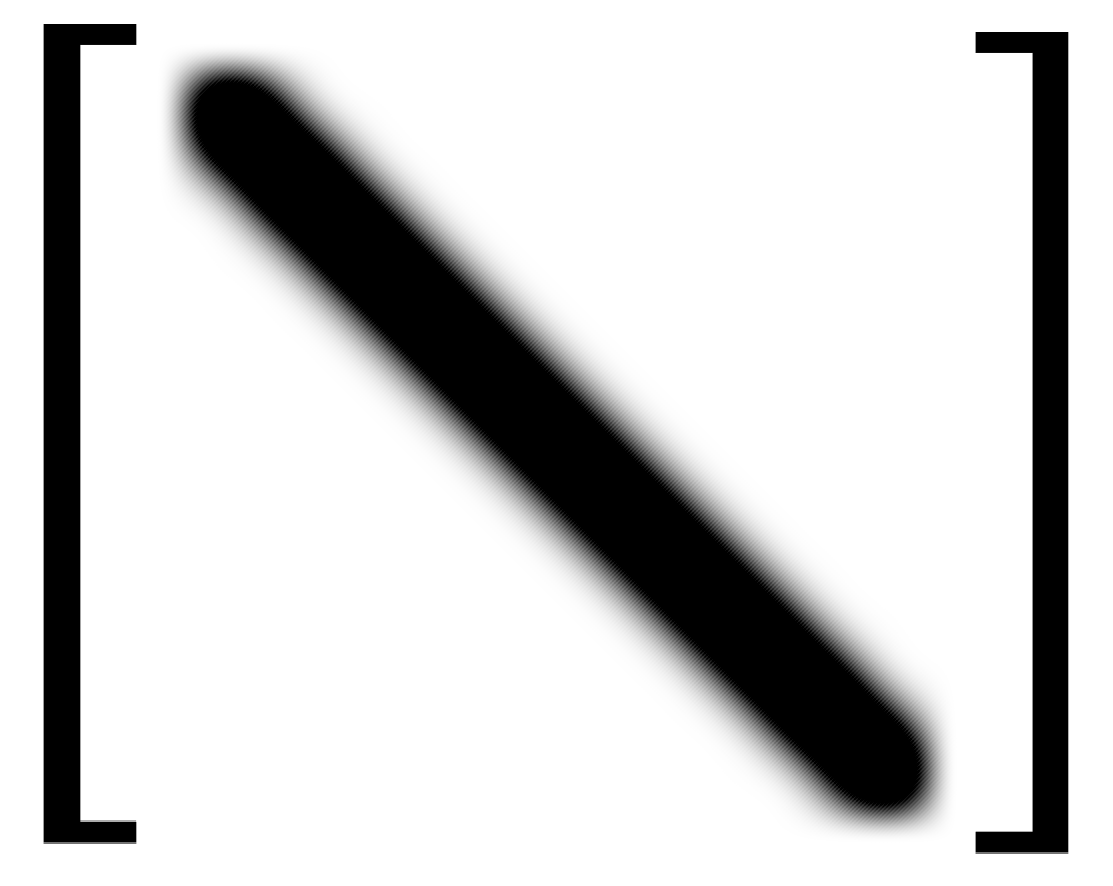}

\vspace{0.5cm}

\includegraphics[scale=0.5]{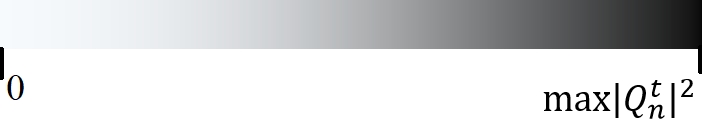}
\end{center}
\caption{Heat maps of matrices $|Q_n^t|^2$ for $n=400$, $y=1$, and $t=\frac{\tau n^2k}{8}$ with $k\in\{ 1,...,8\}$.}
\end{figure}

Figure 2 contains two plots of the entropy $H[|Q_n^t\psi _0|^2/\lVert Q_n^t\psi _0\rVert ^2]$ over time $t$. The left plot shows that the entropy of this system is a roughly periodic function with decaying amplitude. Notice that the most prominent minima occur at approximately $t=\frac{\tau n^2}{8}k$ where $k\in\N$. We deduce that these minima correspond to times at which $Q_n^t\psi _0$ approximates a delta function, as shown in the left side of Figure 4. As $k$ increases, these approximations in some sense lose ``higher frequency" components as they eventually approach the steady-state top eigenvector. However, it is not true that $Q_n^{\tau n^2k/8}$ approximates the identity matrix for all $k$; in fact this approximation holds only for $k$ divisible by 8. Figure 5 shows this sequence of matrices. Notice that at $k=4$, $Q_n^{\tau n^2/2}$ is approximately an anti-diagonal matrix which flips the initial condition. For $k=2,6$, $Q_n^{\tau n^2k/8}$ becomes a weighted sum of the identity approximation and the vector flip approximation. The remaining (odd) values of $k$ result in $Q_n^{\tau n^2k/8}$ becoming a weighted sum of approximations of the identity matrix, the vector flip matrix, a matrix which swaps the top and bottom halves of the vector, and a matrix which flips the top and bottom half of the vector separately.

\begin{figure}
\begin{center}\includegraphics[scale=0.65]{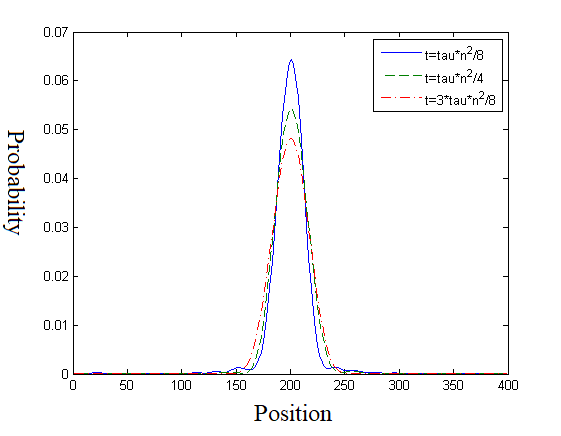}\includegraphics[scale=0.65]{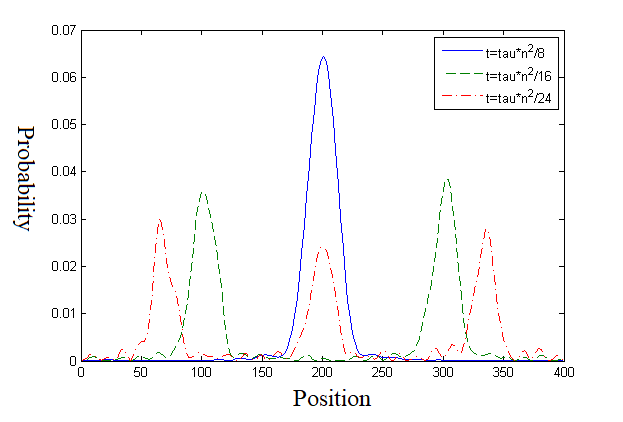}\end{center}

\vspace{0.1cm}

\begin{center}\includegraphics[scale=0.65]{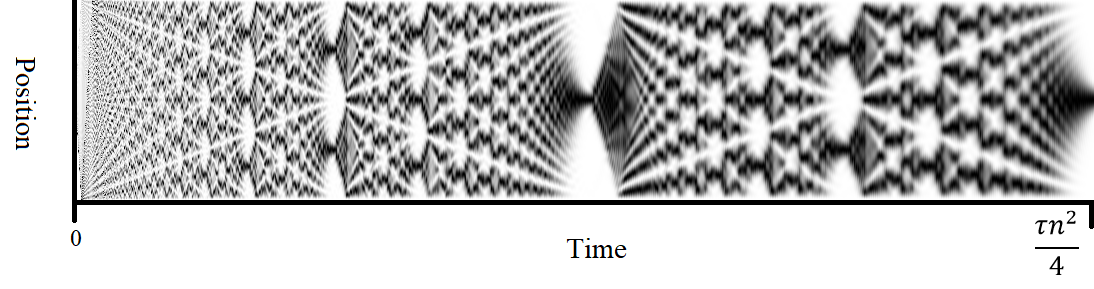}\end{center}

\begin{center}\includegraphics[scale=0.5]{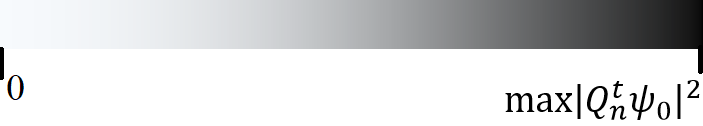}\end{center}
\caption{Plots of $|Q_n^t\psi _0|$ for $n=400$, $y=1$, $\psi _0=|200\rangle |R\rangle$, and various values of $t$.}
\end{figure}

The right plot of Figure 2 shows significant detail at a smaller scale as well. We observe that entropy minima occur approximately at times $t=\frac{\tau n^2p}{8q}$ where $p,q\in\N$ are sufficiently small, in a manner similar to Thomae's function \cite{beanland09}. Moreover, the right plot of Figure 4 suggests that for $\text{gcd}(p,q)=1$, the distribution $|Q_n^t\psi _0|^2$ has $q$ peaks. This observation can be resolved by noting that letting $\tau =4p/\pi y q$ leads to an extra factor of $\exp\left(2\pi ik^2 p/q\right)$ in equations (14) and (17). This causes the eigenvalues to align on a finite set of lines in the complex plane, the number of which is equal to the number of quadratic residues in the ring of integers modulo $q$. By considering the eigenvectors of $Q_n$ from equation (13) as crude approximations of $\sin (\pi kx)$, the mechanism behind these fractional quantum revivals becomes clear. 

\begin{figure}
\begin{center}\includegraphics[scale=0.6]{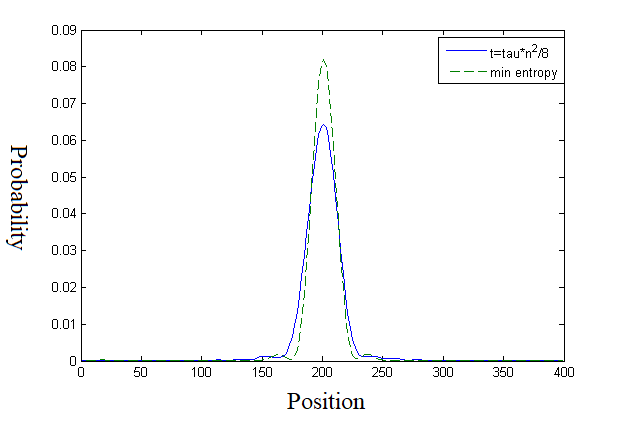}\includegraphics[scale=0.6]{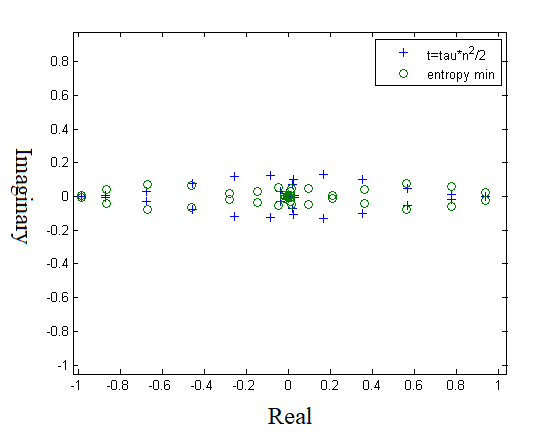}

\vspace{0.5cm}

\includegraphics[scale=0.14]{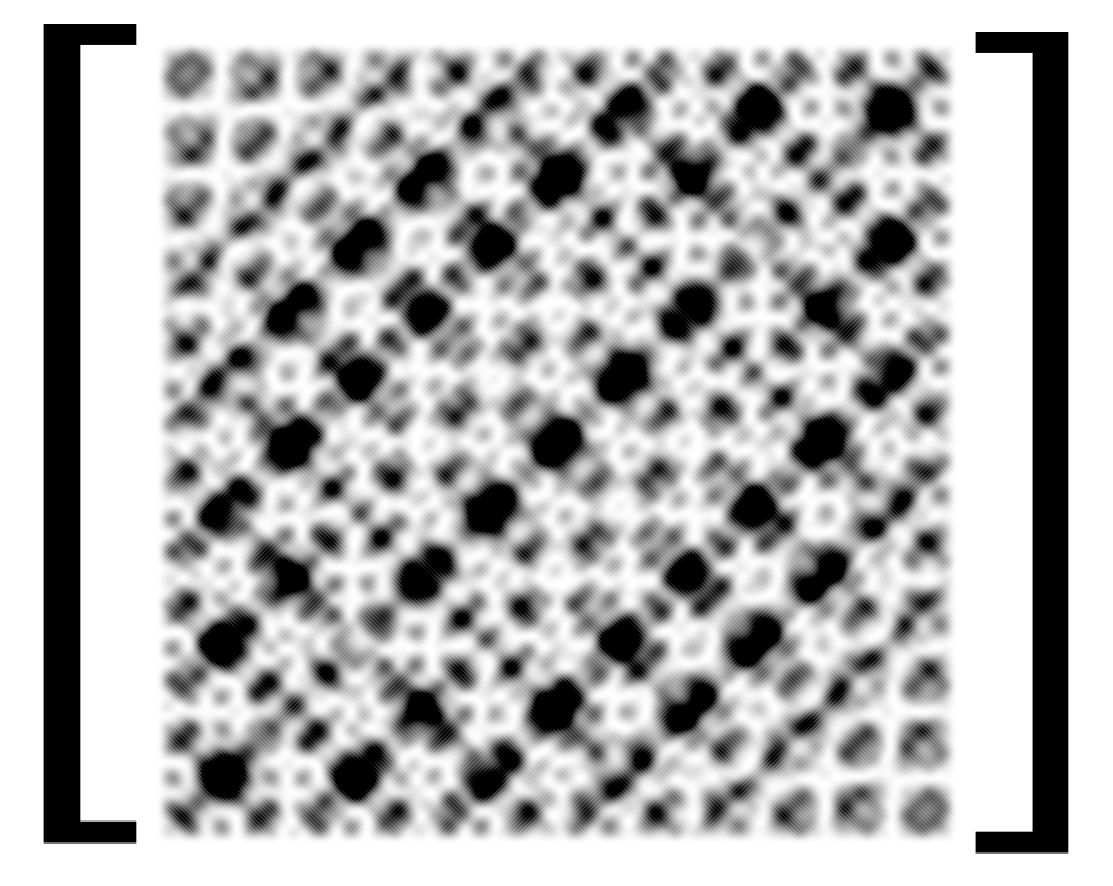}\hspace{0.1cm}\includegraphics[scale=0.14]{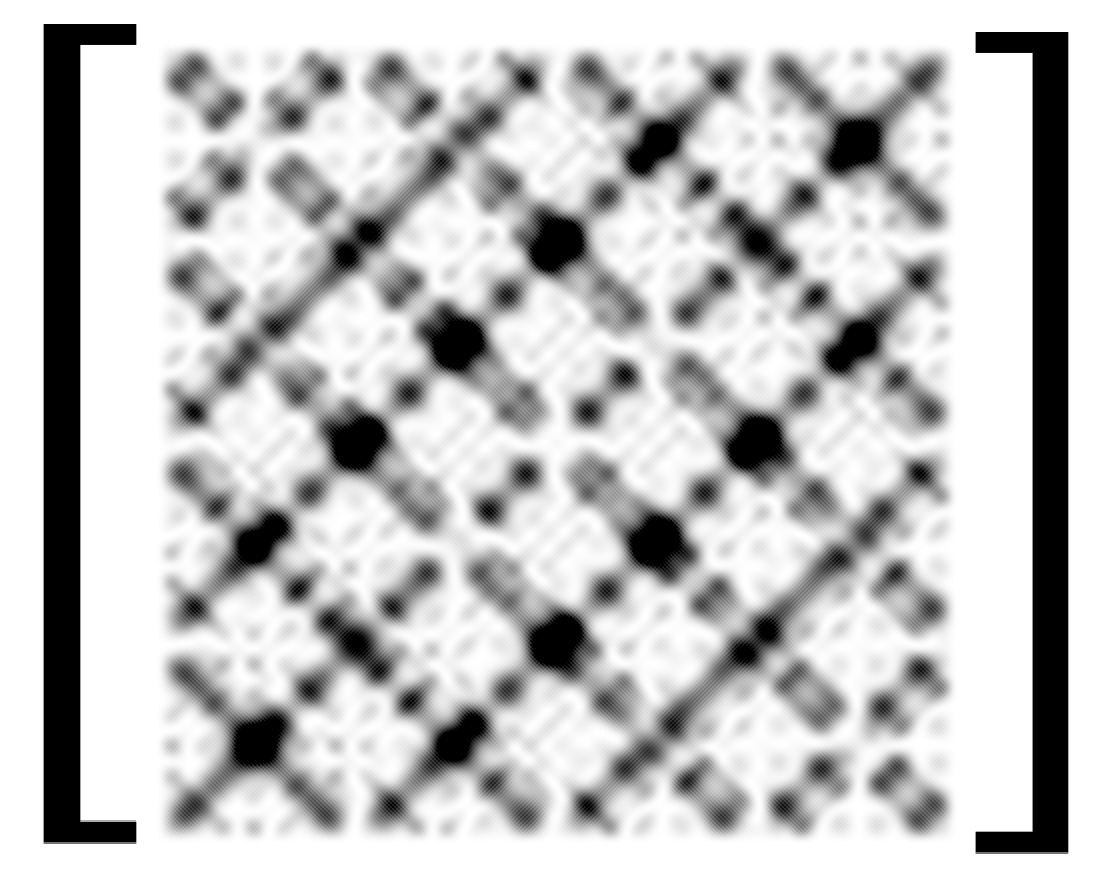}\hspace{0.1cm}\includegraphics[scale=0.14]{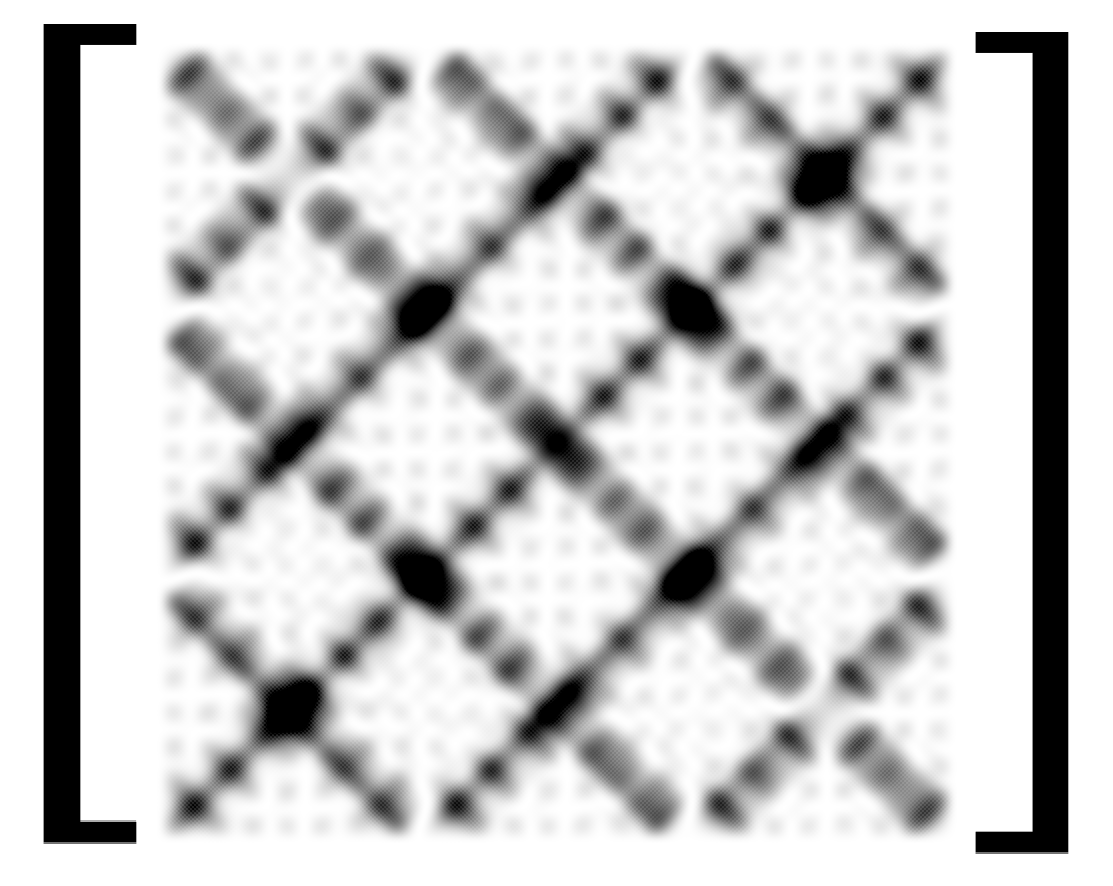}

\vspace{0.5cm}

\includegraphics[scale=0.14]{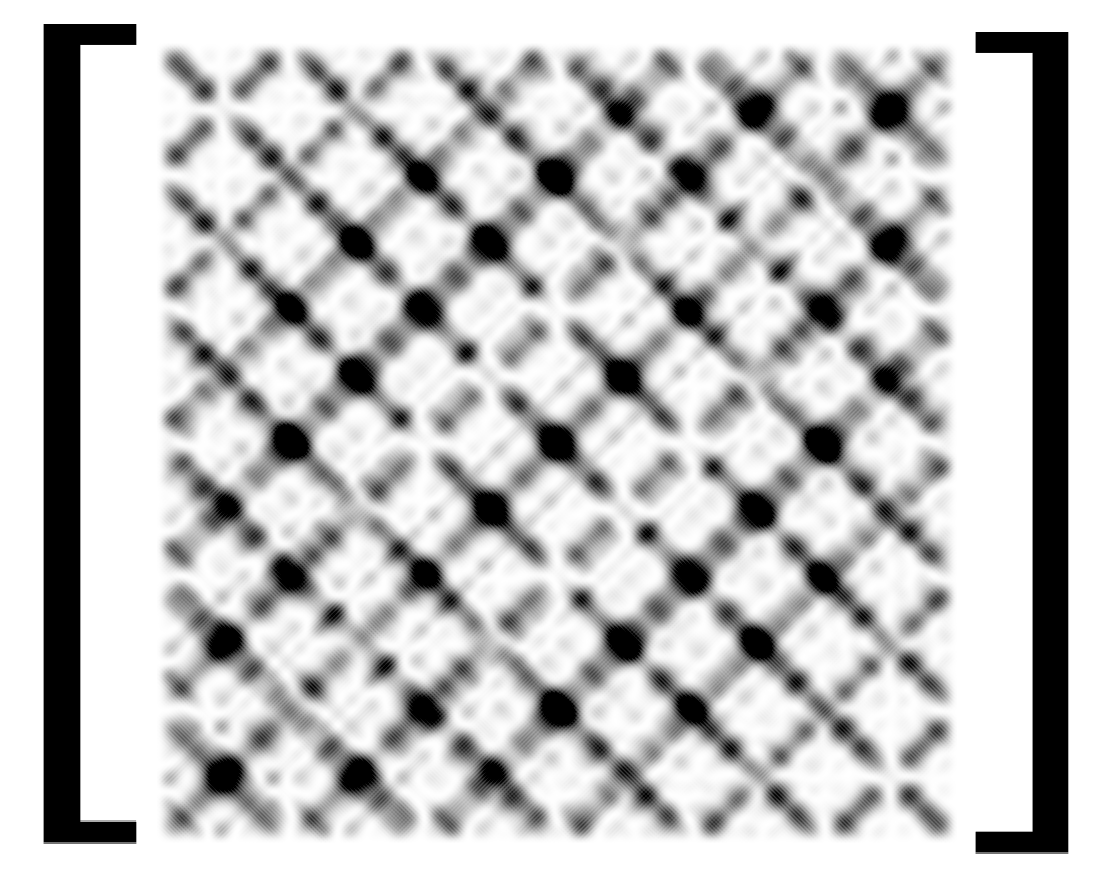}\hspace{0.1cm}\includegraphics[scale=0.14]{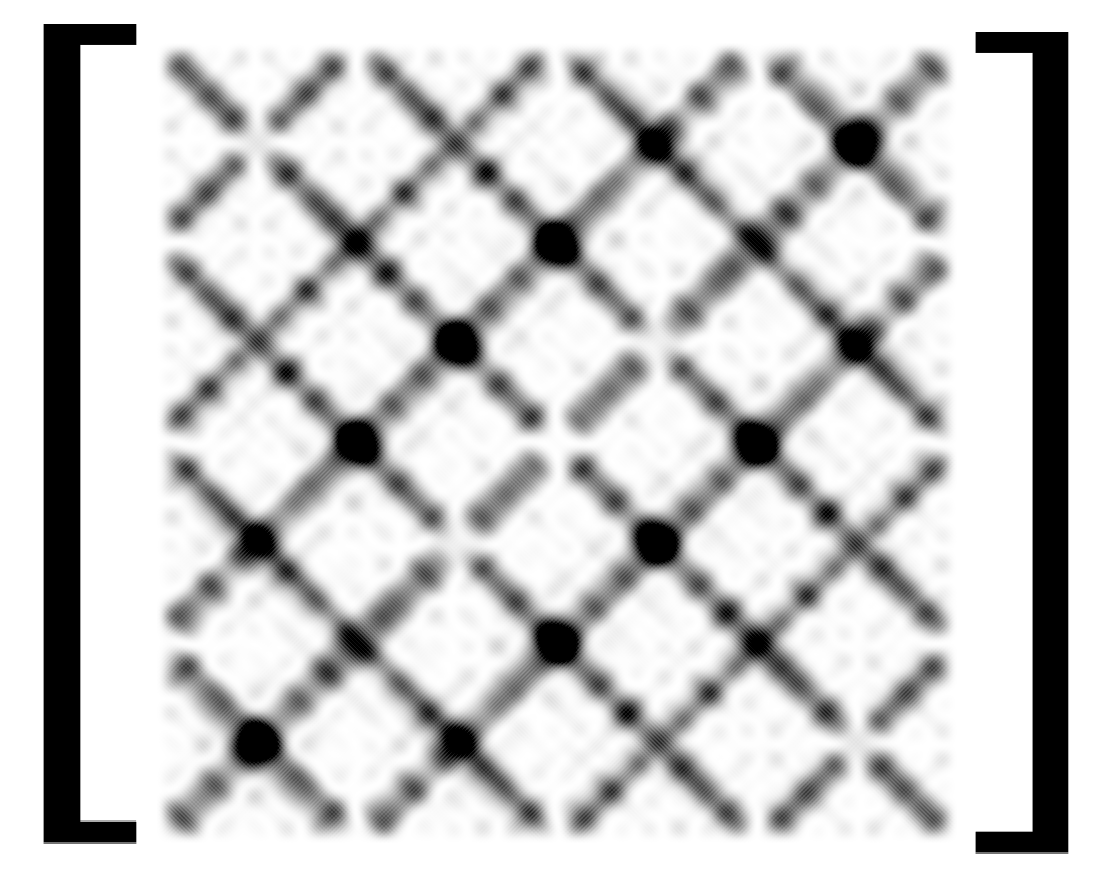}\hspace{0.1cm}\includegraphics[scale=0.14]{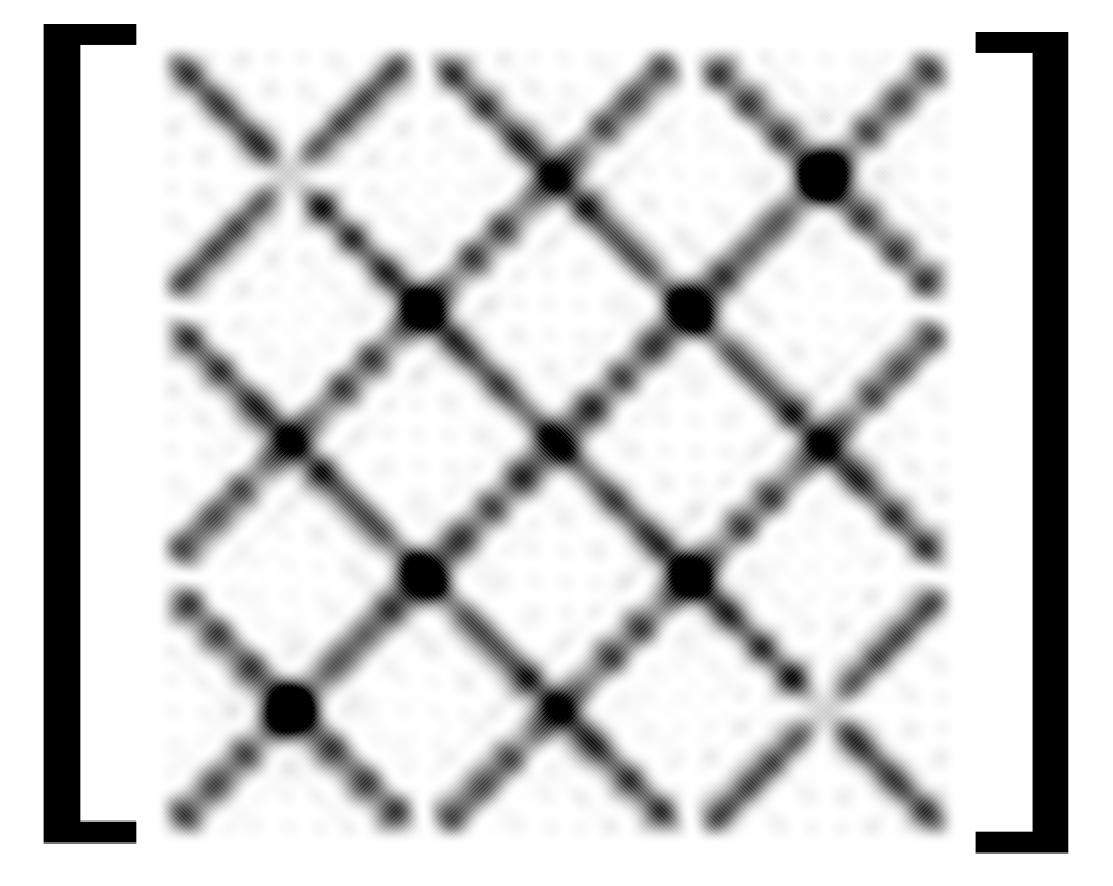}

\vspace{0.5cm}

\includegraphics[scale=0.5]{scale_2.png}
\end{center}

\caption{Illustrations of entropy minimization for absorbing quantum walks with $n=400$ and $y=1$ (\emph{Top Left}) Entropy correction for $|Q_n^t\psi _0|^2$ with $\psi _0=|200\rangle |R\rangle$ and $t=\frac{\tau n^2}{8}=25,465$. Here, the time of minimum entropy is $t=25,627$ (\emph{Top Right}) Entropy corrected eigenvalues of $Q_n^t$ with $t=\frac{\tau n^2}{2}=101,859$. Here, the time of minimum entropy is $t=102,054$.  (\emph{Middle Row}) Heat maps of the matrices $|Q_n^t|^2$ for $t=\frac{\tau n^2}{24}=8,488$, $t=\frac{\tau n^2}{16}=12,732$, and $t=\frac{\tau n^2}{12}=16,977$ respectively. (\emph{Bottom Row}) Entropy corrected plots of the matrices $|Q_n^t|^2$ at times $t=8,603$, $t=12,866$, and $t=17,120$ respectively.}
\end{figure}

However, from the estimate of $(\lambda _{\beta\sqrt{n},n}^\pm e^{\mp i\phi})^{\tau n^2+\rho n}$ in equation (17), we see that the phases of the top $O(\sqrt{n})$ eigenvalues never perfectly align. While letting $\tau =\frac{4}{\pi y}\frac{p}{q}$ and $\rho =0$ gives a good estimate of the time locations of minimum entropy, we can see from Figure 5 that the actual locations of minimum entropy occur at slightly later times as indicated by the improved resolution from the middle to the bottom row of plots. Notice that the eigenvalues at the time of minimum entropy are contained in a narrower band about the real axis than are the eigenvalues at the approximation $t=\tau n^2/2$. In Figure 6, we see various estimates of $\rho$ which lead to entropy minimization. The left plot appears to confirm that $\rho$ converges to a finite value as $n\rightarrow\infty$, while the second plot illustrates how $\rho$ changes as a function of $|a|$. A true estimation of $\rho$ would require an estimate of entropy in the system, and this requires a better estimate of the eigenvectors than has been given. Further still, no easily deducible heuristic gives a usable approximation for $\rho$ as a function of $y$ and $\tau$.

\begin{figure}
\begin{center}\includegraphics[scale=0.65]{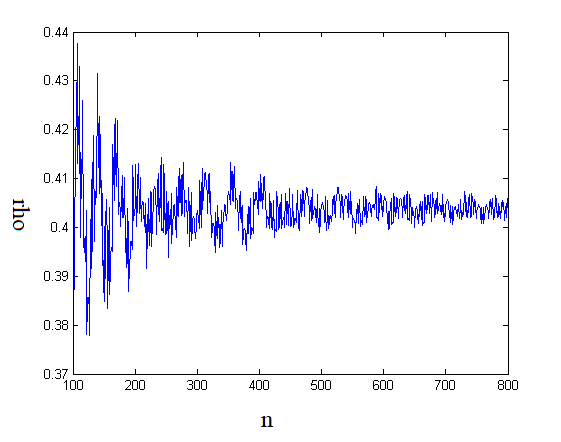}\includegraphics[scale=0.65]{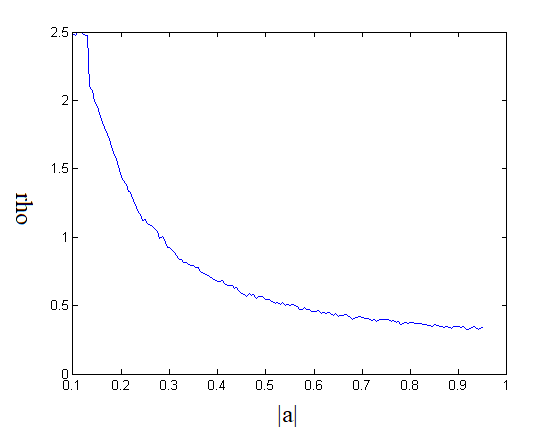}\end{center}
\caption{(\emph{Left}) Estimate of $\rho$ as a function of $n$ for $t=\tau n^2/8$ and $y=1$. (\emph{Right}) Estimate of $\rho$ as a function of $|a|$ with $n=400$ and $t=\tau n^2/8$.}
\end{figure}

These fractional quantum revivals are not unique to the absorbing quantum walk and have manifested in a variety of other quantum systems; the bottom plot in Figure 4 is a fractal that is referred to as a quantum carpet \cite{berry01} \cite{freisch00} \cite{marzoli98}, and was first observed by Henry Talbot in 1836 in the context of optical science \cite{talbot36}. In the case of the particle in an infinite potential well, the spacing of energies arising from the Schr\:{o}dinger equation precisely scales at $k^2$, leading to exact revivals. In the absorbing quantum walk operator $Q_n$ this spacing is not exact, particularly at higher energies. However, only the top $O(\sqrt{n})$ eigenstates contribute to the conditional probability distribution at times $t=O(n^2)$, and the spacing among these eigenvalues is such that approximate revivals are obtained, though the fidelity of these revivals worsens over time until the stable $t=O(n^3)$ regime is reached. This argument breaks down for several purely unitary quantum walks where all eigenstates are relevant. Here, the eigenvalues irrationally wind around the unit circle such that approximate revivals occur only after extremely long times scaling with the size of the lattice; approximate revivals of continuous time quantum walks on the cycle have only been observed for very small lattice sizes. \cite{chandrashekar10}

\section{Extension to Two-Dimensional Absorbing Grover Walk}

As seen in the previous section, fractional quantum revivals arise because the top eigenvalues of $Q_n$ have regular spacing in phase, and the remaining eigenvalues decay to zero exponentially as $t=O(n^{2})$. It should not be surprising that other sufficiently symmetric absorbing quantum walk systems also share this property. For example, consider the two-dimensional absorbing Grover walk operator $Q_{x,y}=\Pi _\text{no}^{B_{x,y}}Q\leftrightarrow (\Z ^2,C_2,G_4,B_{x,y})$. Here, $C_2=\{ (0,1),(0,-1),(1,0),(-1,0)\}$ represents the cardinal directions in $\Z ^2$, $G_n=\frac{2}{n}{\bf 1}_n-I_n$ where ${\bf 1}_n$ is the $n\times n$ matrix filled with ones, and $B_{x,y}=\{ (a,b):a=1\text{ or }a=x,b\in\Z\}\cup\{ (a,b):a\in\Z ,b=1\text{ or }b=y\}$ forms an absorbing ``box". The corresponding $4xy\times 4xy$ operator matrix requires tensor methods to decompose and we postpone this analysis to a later study. From \cite{inui04} \cite{kuklinski17}, localized eigenvectors with eigenvalue $\lambda =1$ take nonzero values on $2\times 2$ areas in the interior of $B_{x,y}$. Since these eigenvectors are the only eigenvectors of $Q_{x,y}$ with eigenvalues $|\lambda |=1$, from Proposition 2.2 in Kuklinski \cite{kuklinski18_2} an initial condition $\phi _0$ which is orthogonal to each of these eigenvectors will eventually decay in norm to zero under $Q_{x,y}^t$. If we let $x=y$ and $\phi _0$ be one such orthogonal initial conditions localized to $(\frac{x}{2},\frac{x}{2})$, then one of the stable distributions in Figure 7 are eventually achieved. These stable distributions correspond to the non-localized top eigenvectors of $Q_{x,y}$.

\begin{figure}
\begin{center}\includegraphics[scale=1.2]{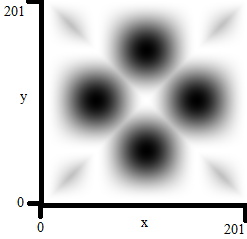}\hspace{1.0cm}\includegraphics[scale=1.2]{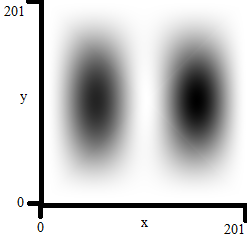}

\vspace{1.0cm}

\includegraphics[scale=1.2]{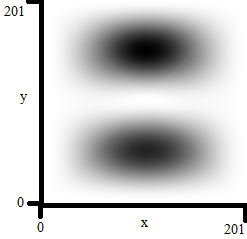}\hspace{1.0cm}\includegraphics[scale=1.2]{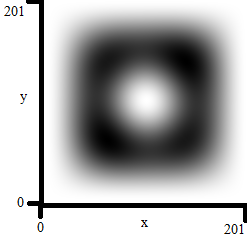}

\vspace{1.0cm}

\includegraphics[scale=0.7]{scale_1.png}
\end{center}
\caption{Non-localized stable distributions of absorbing Grover walk operator $Q_{n,n}$ with $n=201$ and initial conditions localized to $(101,101)$.}
\end{figure}

We document the existence of fractional quantum revivals in the absorbing Grover walk, both for square absorbing boxes and also rectangular absorbing boxes. Figure 8 plots entropy of the systems over time; these graphs depict a similar periodic stratification of entropy minima as the plot in Figure 2, albeit with less regularity. We speculate that these entropy minima occur at times $t=\tau zn^2$ for sufficiently simple $z\in\Q$. Figure 9 displays three of these distributions of the $Q_{200,200}$ absorbing Grover walk at entropy minima; perhaps unsurprisingly the peaks arise in an evenly spaced square grid pattern, although there also appear to be non-negligable diagonal patterns. However for the rectangular $Q_{300,150}$ absorbing Grover walk, the minimum entropy distributions displayed in Figure 10 do not lend themselves so easily to a simple geometric description.

\begin{figure}
\begin{center}\includegraphics[scale=0.65]{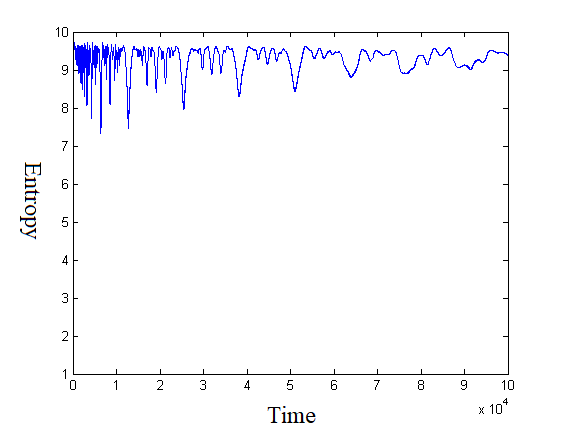}\includegraphics[scale=0.65]{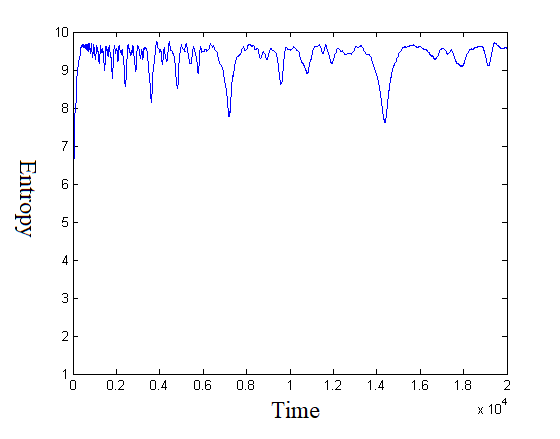}\end{center}
\caption{(\emph{Left}) Plot of entropy over time for $H[|Q_{200,200}^t\psi _0|^2]$ with $\psi _0=|100\rangle |100\rangle |R\rangle$. (\emph{Right}) Plot of entropy over time for $H[|Q_{200,200}^t\psi _0|^2]$ with $\psi _0=|150\rangle |75\rangle |R\rangle$.}
\end{figure}

\begin{figure}
\begin{center}\includegraphics[scale=0.85]{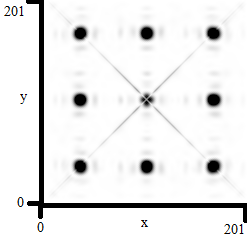}\hspace{0.5cm}\includegraphics[scale=0.85]{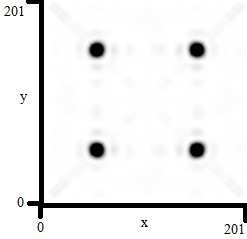}\hspace{0.5cm}\includegraphics[scale=0.85]{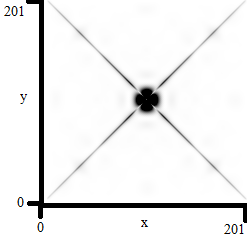}
\end{center}
\caption{Minimum entropy distrubutions of $|Q_{200,200}^t\psi _0|^2$ with $\psi _0=|100\rangle |100\rangle |R\rangle$ (\emph{Left}) $t=4280$ (\emph{Center}) $t=6402$ (\emph{Right}) $t=12752$.}
\end{figure}

\begin{figure}
\begin{center}\includegraphics[scale=0.7]{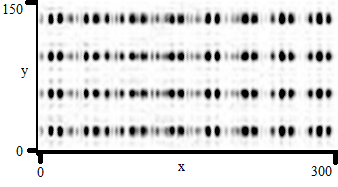}\hspace{0.5cm}\includegraphics[scale=0.7]{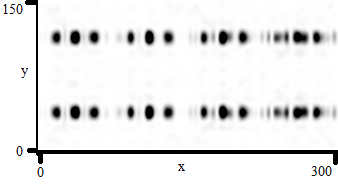}

\vspace{0.5cm}

\includegraphics[scale=0.7]{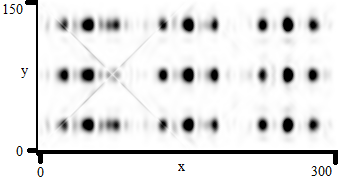}\hspace{0.5cm}\includegraphics[scale=0.7]{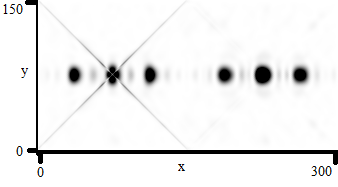}

\vspace{0.5cm}

\includegraphics[scale=0.7]{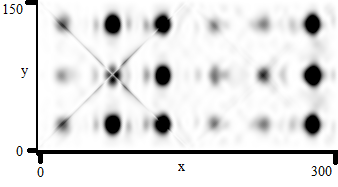}\hspace{0.5cm}\includegraphics[scale=0.7]{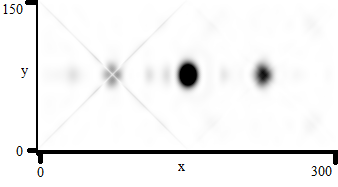}

\vspace{0.5cm}

\includegraphics[scale=0.5]{scale_1.png}
\end{center}
\caption{Minimum entropy distrubutions of $|Q_{300,150}^t\psi _0|^2$ with $\psi _0=|150\rangle |75\rangle |R\rangle$ (\emph{Top Left}) $t=1812$ (\emph{Top Right}) $t=3602$ (\emph{Center Left}) $t=4818$ (\emph{Center Right}) $t=7202$ (\emph{Bottom Left}) $t=9600$ (\emph{Bottom Right}) $t=14374$.}
\end{figure}

\section{Conclusion}

In this paper we have computed eigensystems for one-dimensional finite absorbing quantum walks. The eigenvalues of the corresponding operator $Q_n$ uniformly approach two sectors of the unit circle at $O(n^{-1})$, while the eigenvectors are appoximations of sine waves up to phase. As we consider larger powers $Q_n^t$, we find that the eigenvalues rotate about the origin, and the top $O(\sqrt{n})$ eigenvalues approximately align in the complex plane at regular intervals. This gives rise to fractional quantum revivals described in Section 4. This behavior is found in other sufficiently regular quantum mechanical systems with spacing of eigenvalues proportional to $k^2$.

Several areas of this paper should be expanded upon in future study. If a more robust approximation of the eigenvectors of $Q_n$ are found, one could then make a more informed approximation of entropy for the purpose of entropy minimization. It may also be possible to compute a characteristic polynomial recursion of $Q_{x,y}$, which would then facilitate computation of the stable distributions in Figure 7. Connections to relativistic quantum mechanics could prove fruitful as well; as the quantum walk has been compared to the evolution of a Dirac free particle \cite{bracken_07} \cite{manighalam_19}, one could potentially expand results on Dirac particles in the presence of absorbing boundaries to gather a more general understanding of fractional quantum revivals in dissipative systems \cite{werner_87} \cite{tumulka_16}. Absorbing quantum walks may also find application in scattering theory \cite{komatsu19}. Perhaps an ideal extension of this research would be to find applications of fractional quantum revivals to quantum algorithms. For quantum algorithms involving iterated measurements or absorption, these quantum revivals could provide insight into the behavior of the rest of the quantum state. Furthermore, an absorbing quantum walk could be used imperfectly clone delta potential quantum states \cite{buzek_96} or to purposefully ``low-pass" a quantum state.

\bibliography{biblio}
\bibliographystyle{plain}

\end{document}